\documentclass{aa} 
\usepackage{graphicx,amssymb,amsmath,multirow}
\usepackage[usenames,dvipsnames]{xcolor}
\usepackage{rotating}
\usepackage{txfonts}

\usepackage{lscape}
\usepackage{amsmath}
\usepackage{natbib,twoopt}
\usepackage{longtable}
\usepackage{multicol}
\bibpunct{(}{)}{;}{a}{}{,} 

\newcommand{\bs}{Benchmark Star}
\newcommand{\feh}{[Fe/H]}
\newcommand{\teff}{$\mathrm{T}_{\mathrm{eff}}$}
\newcommand{\logg}{$\log g$}
\newcommand{\vmic}{$v_{\mathrm{mic}}$}
\newcommand{\vmac}{$v_{\mathrm{mac}}$}
\newcommand{\vsini}{$v{\sin{i}}$}
\newcommand{\alfCenA}{$\alpha$~Cen~A}
\newcommand{\alfCenB}{$\alpha$~Cen~B}
\newcommand{\alfCet}{$\alpha$~Cet}
\newcommand{\alfTau}{\object{$\alpha$~Tau}}
\newcommand{\betAra}{\object{$\beta$~Ara}}
\newcommand{\betGem}{\object{$\beta$~Gem}}
\newcommand{\betHyi}{\object{$\beta$~Hyi}}
\newcommand{\betVir}{\object{$\beta$~Vir}}
\newcommand{\delEri}{\object{$\delta$~Eri}}
\newcommand{\epsEri}{\object{$\epsilon$~Eri}}
\newcommand{\epsFor}{\object{$\epsilon$~For}}
\newcommand{\epsVir}{\object{$\epsilon$~Vir}}
\newcommand{\etaBoo}{\object{$\eta$~Boo}}
\newcommand{\gamSge}{\object{$\gamma$~Sge}}
\newcommand{\ksiHya}{\object{$\xi$~Hya}}
\newcommand{\muAra}{\object{$\mu$~Ara}}
\newcommand{\muCas}{\object{$\mu$~Cas}}
\newcommand{\muLeo}{\object{$\mu$~Leo}}
\newcommand{\psiPhe}{\object{$\psi$~Phe}}
\newcommand{\tauCet}{\object{$\tau$~Cet}}
\newcommand{\cygA}{\object{61~Cyg~A}}
\newcommand{\cygB}{\object{61~Cyg~B}}
\newcommand{\Gmb}{\object{Gmb~1830}}
\newcommand{\Arcturus}{\object{Arcturus}}
\newcommand{\Sco}{\object{18 Sco}}
\newcommand{\Procyon}{\object{Procyon}}

\newcommand{\tab}[1]{Tab.~\ref{#1}}
\newcommand{\fig}[1]{Fig.~\ref{#1}}
\newcommand{\sect}[1]{Sect.~\ref{#1}}
\newcommand{\FeI}{$\ion{Fe}{i}$}
\newcommand{\FeII}{$\ion{Fe}{ii}$}

\begin{document}

\title{{Gaia FGK Benchmark Stars - Metallicity}
\thanks{Based on NARVAL and HARPS data obtained within the Gaia DPAC (Data Processing and
Analysis Consortium) and coordinated by the GBOG (Ground-Based Observations for Gaia) working group, and on data retrieved from the ESO-ADP database}
\thanks{The tables are available in electronic form at the CDS via anonymous ftp to cdsarc.u-strasbg.fr 
or via http://cdsweb.u-strasbg.fr/cgi-bin/qcat?J/A+A/}
}

\author{
	P.~Jofr\'e \inst{\ref{IoA},\ref{LAB}}
	\and U.~Heiter \inst{\ref{uppsala}} 
	\and C.~Soubiran \inst{\ref{LAB}} 
	\and S.~Blanco-Cuaresma \inst{\ref{LAB}} 
	\and C.~C.~Worley \inst{,\ref{IoA},\ref{OCA}}
	\and E.~Pancino \inst{\ref{bol1},\ref{bol2}}
	\and M.~Bergemann \inst{\ref{IoA},\ref{MPA}}
	\and T.~Cantat-Gaudin \inst{\ref{pad1},\ref{pad2}}
	\and J.I.~Gonz\'alez~Hern\'andez\inst{\ref{tenerife}}
	\and V.~Hill \inst{\ref{OCA}}
	\and C.~Lardo \inst{\ref{bol1}}
	\and P.~de~Laverny \inst{\ref{OCA}}
	\and K.~Lind \inst{\ref{IoA}}
	\and L.~Magrini \inst{\ref{Arcetri}} 
	\and  T.~Masseron\inst{\ref{IoA},\ref{ULB}}
	\and D.~Montes\inst{\ref{UCM}}
	\and A.~Mucciarelli \inst{\ref{bol3}}
	\and T.~Nordlander \inst{\ref{uppsala}}
	\and A.~Recio~Blanco \inst{\ref{OCA}}
	\and J.~Sobeck\inst{\ref{USA}}
	\and R.~Sordo\inst{\ref{pad1}}
	\and S.~G.~Sousa \inst{\ref{porto}}
	\and H.~Tabernero\inst{\ref{UCM}}
	\and A.~Vallenari \inst{\ref{pad1}} 
	\and S.~Van~Eck\inst{\ref{ULB}}
		 }

\offprints{ \\
P. Jofr\'e, \email{pjofre@ast.cam.ac.uk}; \\ U. Heiter, \email{ulrike.heiter@physics.uu.se}}

\institute{
	{Institute of Astronomy, University of Cambridge, Madingley Rd, Cambridge, CB3 0HA, U.K.}  \label{IoA} 
	\and {LAB UMR 5804, Univ. Bordeaux -- CNRS, 33270 Floirac, France} \label{LAB}
	\and {Department of Physics and Astronomy,  Uppsala University, Box 516, 75120 Uppsala, Sweden} \label{uppsala} 
	\and {Laboratoire Lagrange (UMR7293), Univ. Nice Sophia Antipolis, CNRS, Observatoire de la C\^ote d'Azur,  06304 Nice, France} \label{OCA}
	\and INAF-Osservatorio Astronomico di Bologna, Via Ranzani 1, 40127 Bologna, Italy  \label{bol1} 
	\and {ASI Science Data Center, via del Politecnico s/n, 00133 Roma, Italy} \label{bol2}
	\and Max-Planck-Institut f\"ur Astrophysik, Karl-Schwarzschild-Str. 1, 85741 Garching, Germany \label{MPA} 
	\and {INAF, Osservatorio Astronomico di Padova, Vicolo Osservatorio 5 Padova, 35122 Italy} \label{pad1}
	\and {Dipartimento di Fisica e Astronomia, Universit\`a di Padova Padova, vicolo Osservatorio 3, 35122 Padova, Italy} \label{pad2}
	\and {Instituto de Astrof\' isica de Canarias, 38200 La Laguna, Tenerife, Spain}\label{tenerife}
	\and {INAF/Osservatorio Astrofisico di Arcetri, Largo Enrico Fermi 5 50125 Firenze, Italy}\label{Arcetri}
	\and {Institut d'Astronomie et d'Astrophysique, U. Libre de Bruxelles, CP 226, Boulevard du Triomphe, 1050 Bruxelles, Belgium}\label{ULB}
	\and {Dpto. Astrof\' isica, Facultad de CC. F\' isicas, Universidad Complutense de Madrid, E-28040 Madrid, Spain} \label{UCM}
	\and {Dipartimento di Fisica \& Astronomia, Universit\'a  degli Studi di Bologna, Viale Berti Pichat 6/2, 40127 Bologna, Italy}\label{bol3}
	\and {Department of Astronomy \& Astrophysics, University of Chicago, Chicago, IL 60637, USA} \label{USA}
	\and {Centro de Astrof\'isica, Universidade do Porto, Rua das Estrelas, 4150-762 Porto, Portugal}\label{porto}
		}

\authorrunning{Jofr\'e et. al.}
\titlerunning{Gaia Benchmark Stars metallicity}
   \date{}

\abstract 
   {To calibrate automatic pipelines that determine atmospheric parameters of stars, one needs a sample of stars -- ``benchmark stars'' -- with well defined parameters to be used as a reference.} 
  {We provide a detailed documentation of the determination of the iron abundance of the 34 FGK-type benchmark stars selected to be the pillars for calibration of the one billion Gaia  stars. They cover a wide range of temperatures, surface gravities and metallicities.}
   {Up to seven different methods were used to analyse an observed spectral library of high resolution and high signal-to-noise ratio. The metallicity was determined assuming a  value of effective temperature and surface gravity obtained from fundamental relations, i.e. these parameters were known a priori independently from the spectra.}
   {We present a set of metallicity values obtained in a homogeneous way for our sample of  Benchmark Stars. In addition to this value, we provide a detailed documentation of the associated uncertainties.  Finally, we report for the first time a value of the metallicity of the cool giant \psiPhe. }
  {}


\maketitle

\section{Introduction}
Unlike in the field of photometry or radial velocities, stellar spectral analyses have  up until now lacked a clearly defined set of standard stars spanning a wide range of atmospheric parameters. The \object{Sun} has always been the single common reference point for spectroscopic studies of FGK-type stars. The estimate of  stellar parameters and abundances by spectroscopy is affected by inaccuracies in the input data, as well as  by assumptions made in the model atmospheres and by the analysis method itself. This lack of reference stars, other than the \object{Sun}, makes it very difficult to validate and homogenize a given method over a larger parameter space  \citep[e.g.][]{2008AJ....136.2022L, 2008AJ....136.2050L, 2008AJ....136.2070A, 2010A&A...517A..57J, 2008AJ....136..421Z, 2011AJ....141..187S}.

This is particularly important for the many Galactic surveys of stellar spectra under development \citep[RAVE,][]{2006AJ....132.1645S}; \citep[LAMOST,][]{2006ChJAA...6..265Z}; \citep[APOGEE, ][]{2008AN....329.1018A};  \citep[HERMES,][]{2010gama.conf..319F}; \citep[Gaia,][]{2001A&A...369..339P}; \citep[Gaia-ESO,][]{2012Msngr.147...25G}. Each of these surveys  has developed its own processing pipeline for the determination of atmospheric parameters and abundances, but the different methodologies may lead to a non uniformity of the parameter scales. This is particularly problematic for the metallicities and chemical abundances, which are  important for Galactic studies performed via star counts. It is thus necessary to define a common and homogeneous scale in order to link different spectroscopic surveys probing every part of the Galaxy.

Kinematical and chemical analyses have been used  to study the Milky Way for over a century \citep[e.g.][]{1920ApJ....52...23K,1989ARA&A..27..555G, 2012ARA&A..50..251I}, providing, for example, the  evidence of the existence of the Galactic thick disk \citep{1983MNRAS.202.1025G}. This population contains stars which have different spatial  velocities \citep[e.g.][]{1993A&A...274..181S, 2003A&A...398..141S}, different chemical abundance patterns \citep{2004A&A...415..155B, 2007A&A...465..271R} and ages \citep[for example the works of ][]{1998A&A...338..161F, 2006ApJ...636..804A},  than the thin disk stars. Similarly, much of our knowledge about the Milky Way halo comes from these kind of studies \citep[see review of][]{2008A&ARv..15..145H}. A halo dichotomy similar to that of the disk has been the subject of discussion \citep{2007Natur.450.1020C, 2011MNRAS.415.3807S, 2012ApJ...746...34B}, where the outer halo has a net retrograde rotation and is metal-poor, contrary to the inner halo, which is slightly more metal-rich.  Moreover, the inner halo is composed mainly of old stars \citep[e.g.][]{2011A&A...533A..59J}, although a number of young stars can be observed. The latter may be the remnants of  later accretion of external galaxies. Evidence for these remnants have been found in stellar surveys like by  \citet{2006ApJ...642L.137B}. \citet{2012A&A...538A..21S} found two chemical patterns in nearby halo stars and claim that they have an age difference, supporting the halo dichotomy scenario. 

The analyses of stellar survey data are thus a crucial contribution to the understanding of our Galaxy. The problem arises when one wants to quantify the differences e.g. in  chemical evolution and time of formation of all Galactic components, which are needed to understand the Milky Way as a unique body. A major obstacle in solving this problem is that each study, like those mentioned above, choose their own data sets and methods.  Homogeneous stellar parameters are therefore a fundamental cornerstone with which to put the different Galactic structures in context. The iron abundance (\feh) is of particular importance because it is a key ingredient for the study of  the chemical evolution of stellar systems. Relations between the elemental abundance ratios [X/Fe] versus \feh, where X is the abundance of the element X, are generally used as tracers for the chemical evolution of galaxies \citep[e.g.][ to name a few]{1997ApJ...477..765C, 1998MNRAS.299..535P,2003MNRAS.340..304R, 2009ARA&A..47..371T, 2012A&A...545A..32A, 2013A&A...554A..44A}. Thus, a good determination of the iron abundance is of fundamental importance. 
 
 A major contribution in the study of the Milky Way is expected from the Gaia mission \citep{2001A&A...369..339P}. In particular, the Gaia astrophysical parameters inference system   \citep[Apsis, ][]{2013A&A...559A..74B} will estimate atmospheric parameters of one billion stars. The calibration of Apsis relies on several levels of reference stars, the  first one being defined by  benchmark stars. Some of these stars were chosen to cover the different spectral classifications and to have physical properties known independently of spectroscopy. 
This has motivated us to search for  stars of different FGK types, which we call  Gaia FGK \bs s. Knowing their radius, bolometric flux and distance allows us to measure their effective temperature directly from the Stefan-Boltzmann relation and their surface gravity from Newton's law of gravity.   Our sample of \bs s consists of 34 stars covering different regions of the Hertzsprung-Russell Diagram, representing thereby the different stellar populations of our Galaxy.  It is important to make the comment that our set of FGK \bs s includes some M giant stars. We have decided to include them in the complete analysis described in this paper because we have been successful in analysing them with our methods in a consistent way with respect to rest of the FGK stars of our benchmark sample. However, they should be treated with caution as benchmarks  for FGK population studies.   

In Heiter et al. (2013, in preparation, hereafter Paper~I), we describe our selection criteria and the determination of the ``direct'' effective temperature and surface gravity. In Blanco-Cuaresma et al. (2013, hereafter Paper~II), we present our spectral data of these \bs s and how we treat the spectra in order to build spectral libraries. This article describes the determination of the metallicity using a  library of \bs s compatible with the  pipelines developed for the parameter estimation of the UVES targets from the Gaia-ESO public spectroscopic survey.  For this purpose, up to seven different methods were employed to perform this spectral analysis,  that span from methods using  equivalent widths to synthetic spectra.  Since the aim of this work is to provide a metallicity scale based on the fundamental \teff\ and \logg, we  homogenized our methods by using common observations, atmospheric models and atomic data.

Although the direct application of the reference metallicity is for the homogenization and the evaluation of the different parameter determination pipelines from the Gaia-ESO Survey and the calibration of Apsis, the final set of Benchmark Star parameters and their spectral libraries  provides the possibility to calibrate spectroscopic astrophysical parameters for large and diverse samples of stars, such as those collected by HERMES, SDSS, LAMOST and RAVE. 

The structure of the paper is as follows: In Sect.\ref{lit}, we  review  the metallicity values available in the literature for the \bs s. In Sect.\ref{data}, we describe the properties of the spectra, while the methods and analysis structure  are explained in Sect.~\ref{method}. Our results are presented in Sect.~\ref{results} with an extensive discussion on the metallicity determination in Sect.~\ref{errors}. The paper concludes in Sect.~\ref{conclusions}.

\section{The metallicity of \bs s: reviewing the literature}\label{lit}

\begin{figure}
  \resizebox{\hsize}{!}{\includegraphics{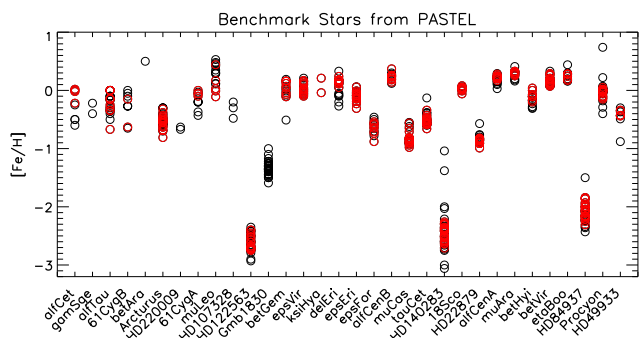}}
  \caption{Spectroscopic metallicities reported for the FGK \bs s in the literature between 1948 to 2012, as retrieved from the PASTEL database \citep{2010A&A...515A.111S}. Black circles: all measurements. Red circles: Only those measurements where \teff\ and \logg\ reported by these work agreed within 100~K and 0.5 ~dex with the fundamental values considered by us (see \tab{table:BMKparams}). }
  \label{fig:feh_pastel}
\end{figure}

The criteria to select the 34 \bs s discussed in this paper can be found in Paper~I. 
Due to their brightness and  proximity, almost every star has previously been studied spectroscopically and has accurate Hipparcos parallax. 
Based on the recently updated PASTEL catalogue \citep{2010A&A...515A.111S}, metallicity values have been reported in 
259 different works until 2012, varying from 
57 [Fe/H] measurements in the case of \object{HD140283} 
to only one measurement for \betAra\ \citep{1979ApJ...232..797L}, and no measurement at all for \psiPhe. 
Figure~\ref{fig:feh_pastel} shows those metallicity values taken from
PASTEL for each Benchmark Star, where in
black color we show all metallicities and in red color only those
where  the \teff\ and \logg\ values agree within 100~K and 0.5~dex, respectively,
with the   respective values  adopted in Paper~I. Note that the \object{Sun} and \psiPhe\
are not included in \fig{fig:feh_pastel} because they are not in PASTEL. 

Recent studies that have analyzed at least 10 Benchmark Stars
are  \citet{2004A&A...420..183A}, \citet[hereafter VF05]{ VF05}, \citet{Luck, 2006AJ....131.3069L}, \citet[hereafter R07]{2007A&A...465..271R},  \citet{Br10}  and \citet[hereafter W12]{Worley2012}, but none of
them have analyzed the complete sample. The literature value for \feh\ that we adopt is the average of the most recent determinations, after 2000, listed in PASTEL.
Table~\ref{table:BMKparams} gives the mean \feh\ with standard deviation and
number of values considered after 3$\sigma$ clipping of all references found in PASTEL after 2000.  For \betAra\ the reported value is the only one available, by \citet{1979ApJ...232..797L}.

Figure~\ref{fig:feh_pastel} shows how  metallicity varies from reference to reference. It is common to have differences of up to 0.5~dex for one star.  Although the scatter significantly decreases when one considers those works with temperatures and surface gravities that agree with our values, there are still some stars which present $\approx$ 0.5~dex difference in \feh, such as \Arcturus\ and the metal-poor stars \object{HD140283}, \object{HD122563} and \object{HD22879}.  Note that \Gmb, \gamSge~ and \object{HD107328} do not have \teff\ and \logg\ that agree with those of Paper~I. 

The stars  are plotted in  order of increasing temperature,  \alfCet\ being the coldest star and \object{HD49933} the hottest one of our sample. Note that \psiPhe\ is colder than \alfCet, but is not plotted in the figure for the reasons explained above.  Cold stars have more scattered metallicity literature values  than hot stars. This could be caused by the fewer works reporting metallicity for cold stars than for hot stars in PASTEL.

 There are many sources  of uncertainties that can slightly affect the results and ultimately produce such different \feh\ values in the literature. 
The methods of determining \feh\ in the literature are {\it highly inhomogeneous}, as they have been carried out by many groups using different assumptions, methodologies, and  sources of data, some of them briefly explained below. An  extensive discussion of how these different aspects affect the  determined parameters of giant stars can be found in \citet{lebzelter2012}, and for solar-type stars in \citet{2012ApJ...757..161T}. The primary aspects are:
\begin{itemize}
\item \textit{Methods}: The analysis of the observed spectra can be based on equivalent widths \citep[e.g.][R07]{Luck, 2006AJ....131.3069L, 2008A&A...487..373S, tab12} or fitting to synthetic spectra \citep[e.g. from VF05,][]{ Br10}. Other  methods different from equivalent widths or
fitting can be used for deriving \feh, like the parametrisation methods based on projections \citep{2010A&A...517A..57J, Worley2012}. Moreover, each method uses a different approach to find the continuum of the spectra.
\item \textit{Atomic data}: For each method the line list can be  built using atomic data from different sources, i.e. \citet{Br10} and VF05 used the VALD database \citep{1999A&AS..138..119K} whereas R07 
 adopted the values given in the NIST\footnote{http://physics.nist.gov/PhysRefData/ASD/lines\_form.html} database \citep{1996atpc.book.....W}.  There are also methods where the atomic data is adjusted to fit a reference star, typically the \object{Sun} \citep[e.g.][]{2004A&A...415.1153S, 2008A&A...487..373S}
 \item \textit{Observations}: For the same star, different observations are taken and analyzed. For example, \citet{2004A&A...420..183A} and R07 studied spectra from the two-coud\'e instruments \citep{1995PASP..107..251T} at the McDonald Observatory and  from the FEROS instrument \citep{2000SPIE.4008..459K} in La Silla.  VF05 used spectra from the spectrometer HIRES \citep{1994SPIE.2198..362V} at Keck Observatory, UCLES \citep{1990SPIE.1235..562D} at the Siding Spring Observatory and the Hamilton spectrograph \citep{1987PASP...99.1214V} at Lick Observatory.  \citet{Worley2012} used FEROS spectra. These spectra differ in wavelength coverage, resolution, flux calibrations and  signal-to-noise ratios. 
 \item \textit{Atmospheric models}:  MARCS  \citep[][and references therein]{MARCS} and Kurucz  atmosphere models are both used throughout the literature and can produce abundance differences of up to 0.1 dex for identical input parameters  \citep{2004A&A...420..183A,2011A&A...534A..53P}. In addition, some groups have started to use three-dimensional (3D) hydrodynamical atmospheric models which can lead to different stellar parameters compared to when using one-dimensional (1D) hydrostatic models\citep[e.g,][]{2007A&A...469..687C}.
 \item \textit{Solar abundances}: Over the past years, the abundances of the \object{Sun} have been updated and therefore metallicities are provided using different solar abundances. \citet{1993A&A...275..101E}, for example,  considered the solar chemical abundances of \citet{1989GeCoA..53..197A} while \citet{2008A&A...484L..21M} refered to the solar abundances of \citet{2005ASPC..336...25A}.   A change in solar composition affects the atmospheric models and therefore the abundances. 
\item \textit{Non-local thermodynamical equilibrium}:  NLTE effects can have a severe impact on the abundance
determinations, especially for the neutral lines of predominantly
singly-ionized elements, like \FeI\ \citep{1999ApJ...521..753T, 2005ARA&A..43..481A, 2009ARA&A..47..481A}. The effect is
typically larger for metal-poor and giant stars \citep{1999ApJ...521..753T,2012MNRAS.427...27B, 2012MNRAS.427...50L}.   Only a few methods make corrections to the abundances due to these effects \citep[e.g.][]{1999ApJ...521..753T, 2001A&A...370..951M}. 
 \end{itemize}
 
  This work attempts to reduce the inhomogeneities found in the parameters of our sample of stars. This is done by re-estimating the metallicity using the same technique for all stars.




\section{Observational Data}\label{data}

The spectra used in this work have very  high signal-to-noise (SNR) and high resolution. 
Since the Benchmark Stars cover the northern and southern hemisphere, it is not possible to obtain the spectra of the whole sample with one single spectrograph. For that reason we have compiled a spectral library collecting spectra from three different instruments: HARPS, NARVAL and UVES. 

The HARPS spectrograph is mounted on the ESO 3.6m telescope \citep{2003Msngr.114...20M}, and the spectra were reduced by the HARPS Data Reduction Software (version 3.1). The NARVAL spectrograph is located at the 2m Telescope Bernard Lyot \citep[Pic du Midi,][]{NARVAL}. The data from NARVAL were reduced with the Libre-ESpRIT pipeline \citep{donati97}. The UVES spectrograph is hosted by unit telescope 2 of ESO's VLT \citep{2000SPIE.4008..534D}. Two sources for UVES spectra are considered, the Advanced Data Products collection of the ESO Science Archive Facility\footnote{http://archive.eso.org/eso/eso\_archive\_adp.html} 
(reduced by the standard UVES pipeline version 3.2, \citealt{2000Msngr.101...31B}), and  the UVES Paranal Observatory Project UVES-POP library \citep[processed with data reduction tools specifically developed for that library]{2003Msngr.114...10B}. More details of the observations and properties of the original spectra can be found in Paper~II.

To have an homogeneous set of data for the metallicity determination, we have built a  spectral library as described in Paper~II. The spectra have been   corrected to laboratory air wavelengths. The wavelength range has been reduced to the  UVES 580 setup, which is from 476 to 684 nm, with a gap from 577 to 584 nm between the red and the blue CCD.  We have chosen this range because it coincides with the standard UVES setup employed by the Gaia-ESO Survey and our methods are developed to work in that range.  Two libraries of spectra are considered: The first one with $R = 70000$, which is the highest common resolution available in our data, the second  one that retains the original resolution ($R > 70000$), which is different for each spectrum and is indicated in \tab{table:BMKparams}.    Finally, each method decided for itself the best way to identify the continuum.
\begin{table*}[ht!]
\caption{Initial parameters and data information for the \bs s. Column description: \feh LIT corresponds to the mean value of the metallicity obtained by works between 2000 and 2012 as retrieved from  PASTEL  \citep{2010A&A...515A.111S}, where $\sigma$\feh\ is the standard deviation of the mean and N represents the number of works considered for the mean calculation (see \sect{lit}). Effective temperature, surface gravity and their respective uncertainties are determined from fundamental relations as in Paper~I and the rotational velocity \vsini\ is taken from literature, with Ref representing the source of this value. The column Source indicates the instrument used to observe the spectrum in the 70~k library (see \sect{runs}), where N, H, U and U.P denote NARVAL, HARPS, UVES and UVES-POP spectra, respectively.  R and SNR represent the resolving power and averaged signal-to-noise ratio of the spectra of the original library (see \sect{runs}), respectively. For stars repeated in the complete 70~k library (see \sect{runs}) the extra source are indicated in the column labeled as ``extra spectra''.   $(^\ast)$:Two spectra in HARPS are available for this star with different wavelength calibration. ($^{\ast \ast}$): There are many spectra of the \object{Sun} taken from different asteroids for HARPS and NARVAL (See Paper~II for details of the library)}              
\label{table:BMKparams}      
\centering                                      
\small

\begin{tabular}{c | c c c c c c c c c   | c c c c}          
\hline\hline                        

star ID & \feh LIT &$\sigma$\feh& N & \teff & $\sigma$\teff & \logg  & $\sigma$\logg & \vsini & Ref \vsini &Source & R (k) & SNR & extra spectra\\    
\hline                                   
\Sco & 0.03	&0.03&15&5747 & 39&	4.43&0.01  	& 2.2 &Saar &	N & 80 & 380 &H\\
\cygA &-0.20&0.11& 5&	4339 &27&	4.43 & 0.16&		0.0 	&Benz&N& 80 & 360 &--\\
\cygB &-0.27&0.00&2	&4045 &25&	4.53 &0.04&	1.7	 &	Benz &N& 80 &450&--\\
\alfCenA &0.20&0.07&9	&5840 &69&	4.31 &0.02&	1.9 &Br10 &H& 115& 430 & U, H$^\ast$\\
\alfCenB &0.24	&0.04&7&5260 &64&	4.54 &0.02& 	1.0 	&Br10 &H&115&460 &-- \\
\alfCet &	-0.26&0.23&8&3796& 65	&0.91 &0.08&	3.0 &Zama &N&80&300& H, U\\
\alfTau &-0.23	&0.3&15&3927 &	40&1.22 &0.10&	5.0&  	Hekk &	N&80&320 &H \\
\object{Arcturus} &-0.54	&0.04&11&4247 &	37&1.59 &0.04& 3.8 &	 	Hekk &N&80&380& H, U, U.P\\
\betAra &0.5&0.00&1&	4073 &	64&1.01 &	0.13&5.4 &Me02&H&115&240 &--\\
\betGem &0.12&0.06&5&	4858 &60&	2.88 &0.05&	2.0  &	Hekk	&H&115&350& --\\
\betHyi 	&-0.11 &0.08  &6&5873 &45&	3.98 &0.02&	3.3 & Re03	&U.P&80& 650&N, H, U\\
\betVir &  0.13 &0.05 &11&6083 &41&	4.08 &	0.01&2.0 &	Br10 &N&80& 410 &H \\
\delEri &0.13 &0.08& 13&5045 &65&	3.77&0.02 	& 0.7&	 	Br10 &N&80& 350 & H, U, U.P \\
\epsEri &-0.07 &0.05& 17&	5050 &	42&4.60 &0.03&	2.4 	&VF05 &U.P&80&1560 & H, U\\
\epsFor &-0.62 &0.12 & 9&	5069 &78&	3.45 &0.05&	4.2 &	Schr &H&115&310 &  --\\
\epsVir 	&  0.12 &0.03  &3&4983 &	61&2.77& 0.01&	2.0&  	Hekk &	N&80&380 &H\\
\etaBoo 	&  0.25& 0.04 & 9&6105 &28	&3.80  &0.02&	12.7	&Br10  & N&80&430& H\\
\gamSge &	-0.31& 0.09 & 2&3807 &49&	1.05 &0.10&	6.0 &	 	Hekk &N&80&460& --\\
\Gmb &-1.34 &0.08 &17&	4827 &55&	4.60 &0.03 &	0.5 	&VF05 &N&80& 410&--\\
\object{HD107328} &-0.30 &0.00  &1	&4590 &59&	2.20 &0.07 	& 1.9	&Mass&N&80&380& H\\
\object{HD122563} &-2.59 &0.14 & 7	&4608 &60	&1.61 &0.07 	& 5.0& Me06	&N&80&300& H, U, U.P\\
\object{HD140283} &-2.41& 0.10 &10&	5720 &	120&3.67&0.04 &	5.0& 	Me06 &N&80&320& H, U, U.P\\
\object{HD220009} & -0.67 &0.00  &1&	4266 &	54&1.43 &0.10	&1.0& 	Me99 &	N&80&380& --\\
\object{HD22879} &	-0.85 &0.04& 16&5786 &	89&4.23 &0.03	&4.4& Schr &	N&80&300 &--\\
\object{HD49933} &	-0.39 &0.07&  5&6635 &	91&4.21&	0.03&10.0 & Br09 &	H&115&310& --\\
\object{HD84937} &	-2.08 &0.09 &13&6275 &97	&4.11  &0.06&	5.2 	&Me06&H&115&480& N, U, U.P\\
\ksiHya 	& 0.21 &0.00  &1&5044 &38	&2.87 &0.01&	2.4 	&Br10 &H&115&370 &-- \\
\muAra & 0.29 &0.04 &12&	5845 &	66&4.27 &0.02&	2.3 &Br10 &	U&105&420 &\\
\muCas~A &-0.89 &0.04 &14&	5308 &	29&4.41  &0.02&	0.0& 	 	Luck &N&80&280& U\\
\muLeo &	 0.39 &0.10 & 4&4433 &60	&2.50&0.07 	& 5.1 &	 	Hekk &N&80& 400&-- \\
\Procyon &-0.02 &0.04 &18&	6545 &84&	3.99 &0.02&	2.8 & Br10&	U.P&80&760&N, H, U\\
\psiPhe & --& --  & 0&	3472 &	92&0.62 &	0.11&3.0 &	Zama &U&70&220& --\\
\object{Sun} &	0.00 & 0.00 & 0& 5777 &	1&4.43  &2E-4&	1.6 &VF05	&H&115&350& H, N, U$^{\ast \ast}$\\
\tauCet &-0.53 &0.05 &17&	5331& 43	&4.44 &0.02&	1.1& Saar &	N&80&360& H\\
\hline   \hline                                          
\end{tabular}
\normalsize
\tablebib{
(Saar)~\citet{Saar}; (Benz)~\citet{Benz}; (Br10)~\citet{Br10}; (Zama)~\citet{Zama}; (Hekk)~\citet{Hekk}; (Me02)~\citet{Me02}; (Re03)~\citet{Re03}; (VF05)~\citet{VF05}; (Schr)~\citet{Schr}; (Mass)~\citet{Mass}; (Me06)~\citet{Me06}; (Me99)~\citet{Me99}; (Br09)~\citet{Br09}
}
\end{table*}

\section{Method}\label{method}

For consistency, we have used common material and assumptions as much as possible, which are explained below. In this section we also give a brief description of each metallicity determination method considered for this work.

\subsection{Common material and assumptions}

The analysis is based on the principle that the effective temperature and  the surface gravity of each star are known. These values (indicated in  \tab{table:BMKparams}) are obtained independently from the spectra using fundamental methods, i.e., taking the angular diameter and bolometric flux to determine the effective temperature and the distance, angular diameter and mass to determine surface gravity.   
 In our analysis, we fix \teff\ and \logg\ values, as well as
rotational velocity (values also indicated in  \tab{table:BMKparams}). The latter were taken from the literature, for which the source is also indicated in  \tab{table:BMKparams}.  For those methods where a starting value for the metallicity is needed, we set \feh = 0.

We used the line list  that has been prepared for the analysis of the stellar spectra for the Gaia-ESO survey (Heiter et al. 2014, in prep, version 3, hereafter GES-v3). The line list includes simple quality flags like ``yes'' (Y), ``no'' (N) and ``undetermined'' (U). These were assigned from an inspection of  the line profiles and the accuracy of  the $\log gf$ value for each line based on comparisons of synthetic spectra with a spectrum of the \object{Sun} and of \Arcturus.  If the profile of a given line is well reproduced and its $gf$ value is well determined, then the line has `Y/Y''. On the contrary, if the line is not well reproduced (also due to blends) and the $gf$ value is very uncertain, the line is marked with the flag ``N/N''. We considered all lines except those assigned with the flag ``N''  for the atomic data or the line profile. Finally, all methods used the 1D hydrostatic atmosphere models of MARCS \citep{MARCS}, which consider local thermodynamical equilibrium (LTE), and plane-parallel or spherically symmetric geometry for dwarfs and giants, respectively.  These atmospheric models were chosen in order to be consistent with the spectral analysis of the UVES targets from the Gaia-ESO Survey. \\

\subsection{Runs}\label{runs}
Three main analyses were made, as explained below. These runs allow us to study the behavior of our results under different methods, resolutions and instruments.

\begin{enumerate}
\item \textit{Run-nodes}: One spectrum per star at $R=70000$,  where for stars with more than one spectrum available in our library, the ``best'' spectrum was 
selected by visual inspection.  The evaluation was mainly based on the behavior of the continuum, but also considered the SNR and the amount of cosmic ray features and telluric absorption lines. 
The source of the spectra used for this test is indicated in  \tab{table:BMKparams}. Hereafter, we call this set of data the ``70~k library''.  The purpose of this run  was to have a complete analysis and overview of the performance of different methods for  a well-defined set of spectra.

\item  \textit{Run-resolutions}: The same selection of spectra as in \textit{Run-nodes}, but using the original resolution version of the library. This value is indicated in \tab{table:BMKparams}. This run allowed us to make a comparative study of the impact of resolution on the accuracy of the final metallicity. This set of spectra is hereafter called the ``Original library''.

\item \textit{Run-instruments}:  All available spectra obtained with several instruments, convolved to R=70000, i.e. several results for each star. The source of the available spectra for each star (when applicable) is indicated in the last column of \tab{table:BMKparams}.  Hereafter we call this data set the ``complete 70~k library''. 
This run gave us a way to study instrumental effects, and to assess the internal consistency of the metallicity values with regard to the spectra being employed.  
\end{enumerate}

\subsection{Nodes method description}\label{nodes}

In this section we explain the methods considered for this analysis. They vary from fitting synthetic spectra to observed spectra to classical equivalent width (EW) methods.  
Since this analysis is based on 1D hydrostatic atmospheric models, the microturbulence parameter also needed to be taken into account. We considered the value of \vmic\ obtained from the relations of M. Bergemann and V. Hill derived for the analysis of the targets from the Gaia-ESO Survey (hereafter  GES relation). Some of the methods determine this parameter simultaneously with \feh\ using as an initial guess the GES relation, while others kept \vmic\ fixed to the value obtained from the relation. In the following, we will explain briefly each method individually.

\subsubsection{LUMBA}
\indent {\it Code description:} The LUMBA-node (Lund, Uppsala, MPA, Bordeaux, ANU\footnote{ Lund: Lund Observatory, Sweden; Uppsala: Uppsala University, Sweden; MPA: Max-Planck-Institut f\"ur Astrophysik, Germany; Bordeaux: Laboratoire d'Astrophysique de Bordeaux, France; ANU: Australian National University, Australia}) uses the SME \citep[Spectroscopy Made Easy, ][]{sme, VF05}
~code (version 298) to analyse the spectra. This tool performs an automatic parameter optimization using a  chi-square minimization algorithm.
Synthetic spectra are computed by a built-in spectrum synthesis code for a set of global model parameters and spectral line data. A subset of the global parameters is varied to find the parameter set which gives the best agreement between observations and calculations.   In addition to the atmospheric models and line list as input, SME requires masks containing information on the spectral segments that will be analysed, the absorption lines that will be fitted, and the continuum regions which are used for continuum normalisation. The masks have to be chosen so that it is possible to analyse homogeneously the same spectral regions for all stars. To create the masks, we  plotted  the normalised fluxes of all Benchmark Stars and looked for those  lines  and  continuum points that are present in all stars.  The analysis of the LUMBA node was mainly carried out by P. Jofr\'e, U. Heiter, C. Soubiran, S. Blanco-Cuaresma, M. Bergemann and T. Nordlander.

\indent {\it Iron abundance determination:} We made 3 iterations with SME: (i)  determine only metallicity starting from \feh=0 and fixing \vmic~ and macroturbulence velocity (\vmac)  to the values obtained from the GES relations; (ii) determining \vmic~ and \vmac\ fixing the  \feh~value obtained in the previous iteration (see below); (iii)  determination of \feh, including a final correction of { of radial velocity for each line which accounts for residuals in the wavelength calibration or line shifts due to thermal motions \citep{2012A&A...544A.125M}},  using as starting values those obtained in the previous iterations. To validate the ionization balance in our method, we built two sets of masks for \FeI\ and \FeII\ separately.

\indent {\it Broadening parameters:} We estimated the microturbulence and macroturbulence parameters in an additional run with SME. For that, we created a mask including all strong neutral lines with $-2.5 > \log gf > -4.0$ in the spectral range of our data. This value was chosen because lines in this  $\log gf$ regime are sensitive to \vmic~with SME \citep{sme}. To determine the broadening parameters we considered the initial values obtained from the GES relation and fixed with SME \teff\, \logg\ and \feh.

\indent {\it Discussion:} Special treatment was necessary for the metal-poor stars  with $\mathrm{\feh} \leq -0.6$ and for the cold stars with \teff$ \leq 4100$ K. In the case of the metal-poor stars, a significant number of lines from the line masks were not properly detected resulting in the spectra being incorrectly shifted in radial velocity. Since the library is in the laboratory rest frame, we decided not to make a re-adjustment of the radial velocity for these stars.   Cold stars needed a special line mask. In many segments molecular blends were very strong, making it impossible to obtain a good continuum placement and also a good fit between the observed and the synthetic spectra. Moreover, determining iron abundances of blended lines with molecules that are not included in our  line list results in an incorrect estimation of the true iron content  in the atmosphere.  We looked at each spectrum individually and selected the unblended iron lines.

\subsubsection{Nice}
\indent {\it Code description:}  The pipeline is built around the stellar parameterisation algorithm MATISSE (MATrix Inversion for Spectrum SynthEsis) which has been developed at the Observatoire de la C\^{o}te d'Azur primarily for use in Gaia RVS\footnote{Radial Velocity Spectrometer} stellar parameterization pipeline \citep{Recio-Blanco2006}, but also for large scale projects such as AMBRE \citep{Worley2012, deLaverny2012} and the Gaia-ESO Survey. MATISSE simultaneously determines the stellar parameters ($\theta$: \teff, \logg,  [M/H] and [$\alpha$/Fe]\footnote{ The metallicity [M/H] is derived using spectral features of elements heavier than helium while the [$\alpha$/Fe] determination uses spectral features of $\alpha-$elements} of an observed spectrum $O(\lambda)$ by the projection of that spectrum onto a vector function $B_{\theta}(\lambda)$. The $B_{\theta}(\lambda)$ functions are optimal linear combinations of synthetic spectra $S(\lambda)$ within the synthetic spectra grid. For this work,
we adopted the synthetic spectra grid built for the Gaia-ESO survey, by using the {same line list and atmosphere models as the other nodes} and the GES relation for the microturbulence.  A full documentation on how this grid is computed is found in \cite{deLaverny2012}. The analysis done by the Nice group was mainly carried out by C.~C.~Worley, P.~de Laverny,  A.~Recio-Blanco and V.~Hill.

\indent {\it Iron abundance determination:} The wavelength regions selected for this analysis were based on the Fe line mask used by LUMBA. Continuum regions of minimum 8~\AA\ were set about each accepted Fe line or group of lines.

\indent {\it Broadening parameters:} Since this method is restricted to fit synthetic spectra from a pre-computed grid, \vmic\ was determined from the best fit of spectra computed using the GES relation. 

\indent {\it Discussion:} Holding \teff\ and \logg\ constant and allowing metallicity to vary, is not fundamentally possible for MATISSE in the current configuration as MATISSE converges on all the parameters simultaneously.  
MATISSE does accept a first estimate of the parameters, which were set in this case to the fundamental \teff\ and \logg\   and solar [M/H] and [$\alpha$/Fe]. However MATISSE then iterates freely through the solution space to converge on the best fit stellar parameters for each star based on  the synthetic spectra grid.

Additionally a direct comparison of the normalized observed spectrum to the synthetic spectra by $\chi^2$-test was carried out. The synthetic spectra were restricted to the appropriate constant \teff\ and \logg\ with varying [M/H] and [$\alpha$/Fe]. This test did not require the MATISSE algorithm and only provided grid point stellar parameters. However,  it was useful as a confirmation of the MATISSE analysis, and also a true test for which \teff\ and \logg\ could be held constant allowing metallicity to vary. In addition, this is a useful analysis as a  validation of the grid of synthetic spectra available for the Gaia-ESO Survey.

 This configuration  of considering only regions around Fe lines, performed well for metal-rich dwarfs but was more problematic for low gravity and metal-poor star. Three potential reasons are (a) poor representation of the ionization balance due to the small number of \FeII\ lines; b) strong lines were excluded from the regions, the wings of which are typically good gravity indicators; and c) normalization issues for these small spectral regions around Fe lines.  

However, even for the problematic stars, where the $\log g$ $B_{\theta}(\lambda)$ functions did show a lack of strong sensitivity due to a lack of strong features, and the regions of reasonable $\log g$ sensitivity ($\sim$ 5000~\AA\ to 5200~\AA) were difficult to normalize accurately, MATISSE found the solution for each star that best fit this configuration of the synthetic grid. This was confirmed in most cases by the $\chi^2$-test. We remind that the final provided solutions here do not represent those favoured by a full-MATISSE analysis, because of the a-priori fixed \teff\
and \logg\ and the selection of only iron lines in the spectral windows. Some consequences of this fixed analysis for MATISSE
are discussed below.
`
\subsubsection{ULB (Universit\'e Libre de Bruxelles)}

\indent {\it Code description:} The ULB node uses the code BACCHUS (Brussels Automatic Code for Characterising High
accUracy Spectra), which consists of three different modules 
designed to derive abundances, EWs, and stellar parameters.
The current version relies on an interpolation of the grid of atmosphere models
using a thermodynamical structure as explained in \citet{MasseronPhD}. Synthetic spectra are computed using the
radiative transfer code TURBOSPECTRUM \citep{Alvarez1998,Plez2012}. This analysis was carried out mainly by T. Masseron and S. Van Eck.

\indent {\it Iron abundance determination:} The iron abundance determination module includes local continuum placement (adopted from  spectrum synthesis using the full set of lines), cosmic and
telluric rejection algorithms, local SNR estimation, and selection of observed
flux points contributing to the line absorption.  
Abundances are derived by comparison of the observation with a set of
convolved synthetic spectra with different abundances using four different comparison methods: 
$\chi ^2$ fitting, core line intensity, synthetic fit, and EWs. A decision tree is constructed from those methods to select the best matching abundances.

\indent {\it Broadening parameters:} Microturbulence velocity was determined in an iterative way together with the iron abundances. 
For that, a new model atmosphere was taken into account for
the possible change in metallicity by adjusting the microturbulence
velocity. Additionally, a new convolution parameter for the
spectral synthesis encompassing macroturbulence velocity, instrument
resolution of 70000 and stellar rotation was determined and adopted if necessary.

\subsubsection{Bologna}\label{bol}
\indent {\it Code description:} The analysis is based on the measurement of EW. This was done
using DAOSPEC \citep{daospec},  run through DOOp \citep{2013arXiv1312.3676C}, a program that automatically configures some of the
DAOSPEC parameters and makes DAOSPEC run multiple times until the input and
output FWHM\footnote{DAOSPEC uses the same FWHM (scaled with wavelength) for all lines, thus, an input FWHM is required from the user to be able to separate more easily real lines from noise (which generally has a FWHM of 1--2 pixels). Later, the code refines the FWHM and determines the best value from the data, thus producing an output FWHM}   of the absorption lines agree within 3\%. The analysis of the Bologna method was mainly carried out by E.~Pancino, A.~Mucciarelli and C.~Lardo.

\indent {\it Iron abundance determination:} The abundance analysis was carried out with GALA \citep{gala}, an
automatic program for atmospheric parameters and chemical abundances determination from
atomic lines, based on the Kurucz suite of programs \citep{klinux, kurucz}. Discrepant lines with respect to the fits of the slopes of Fe  abundance versus EW, excitation potential, and wavelength were rejected with a 2.5$\sigma$ cut, as well as lines with too small or to large EW (depending on the star). 

\indent {\it Broadening parameters:} We looked for the best \vmic\ whenever possible, by looking for the solution
which minimised the slope of the [Fe/H]~vs.~EW relation.  If for some stars it was not possible to converge to a meaningful value of
\vmic\ (mostly because not enough lines in the saturation regime were measurable
with a sufficiently accurate 
Gaussian fit), we used the GES relations which provided a flat [Fe/H]~vs.~EW relation.

\indent {\it Discussion:}
 Some of the stars, which have deep molecular
bands or heavy line crowding, had to be re-measured with an exceptionally high
 order  in the polynomial fit of the continuum (larger than 30). The stars which
needed a fixed input \vmic\ were: 61~Cyg~A and B, \betAra, \epsEri, and
\Gmb.


\begin{figure*}
  \resizebox{\hsize}{!}{\includegraphics[angle=90]{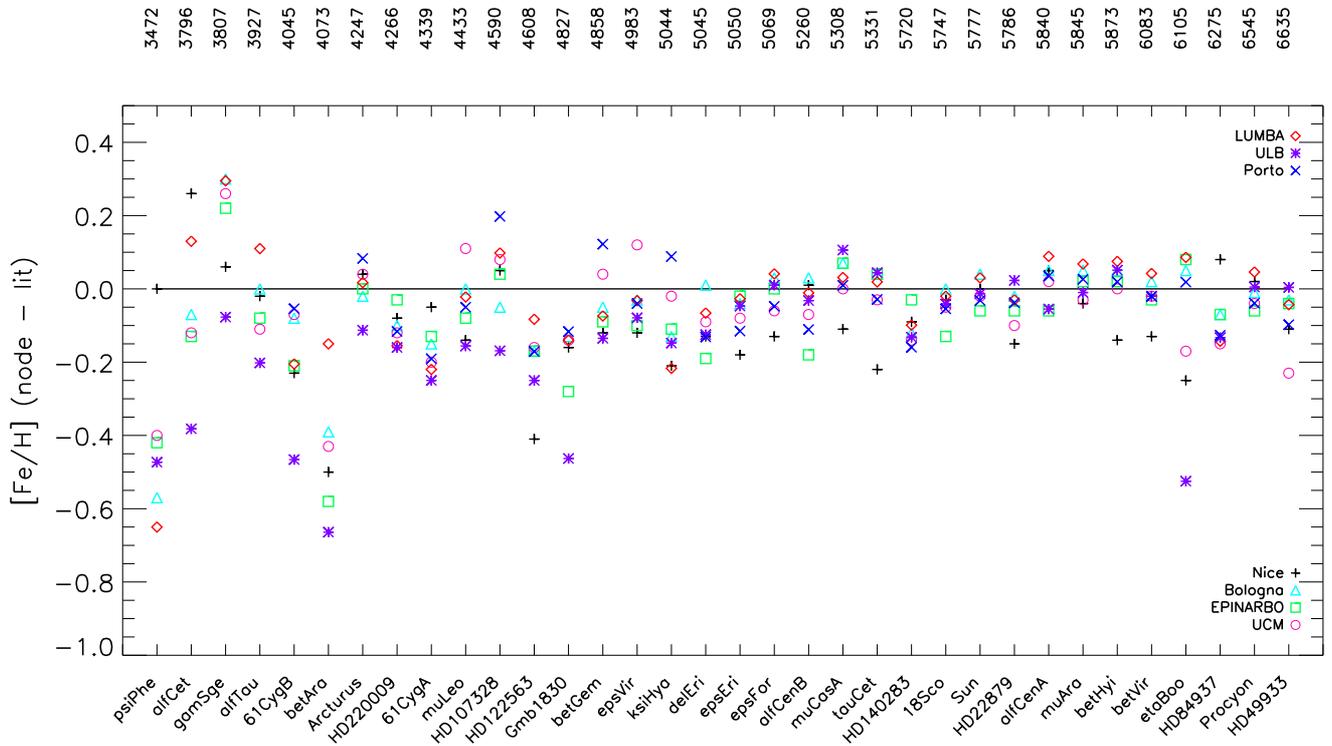}}
  \caption{Difference between the metallicity obtained by each node and the mean literature value (see \sect{lit}). 
  Stars are ordered by effective temperature. Different symbols correspond to the different methods, which are indicated in the legend.}
  \label{fig:run01}
\end{figure*}

 \subsubsection{EPINARBO}
\indent  {\it Code description:}  The EPINARBO-node (ESO-Padova-Indiana-Arcetri-Bologna\footnote{ European Southern Observatory;  Osservatorio Astronomico di Padova, Italy; Indiana University, USA; Osservatorio Astrofisico di Arcetri, Italy; Istituto Nazionale di Astrofisica, Italy.}) adopts  a code, {\sc FAMA} \citep{2013arXiv1307.2367M}, based on
an automatization of MOOG \citep[v.2010]{1973PhDT.......180S}, which is based on EWs determined in the same way as in the Bologna method (see \sect{bol})\footnote{These measurements were carried out
independently from the Bologna ones, with slight differences in the
configuration parameters (continuum  polynomial fit order, input FWHM, starting radial
velocity, and so on), leading to mean differences  that are generally of the order or $\pm$1\%, except for a few stars which could have a mean difference up to $\simeq$3\%}. The analysis of this node was mainly carried out by T.~Cantat-Gaudin, L.~Magrini, A.~Vallenari and R.~Sordo.

\indent {\it Iron abundance determination:} For the purpose of determination of metallicity only, we  fixed the
effective temperature and surface gravity, and computed \vmic\ with the adopted formulas of the GES relation.  In this way, by keeping  these three
atmospheric parameters fixed, we  obtained the average of both  neutral and ionised iron abundances, discarding with one-$\sigma$ clipping those abundances which are discrepant.  

\indent {\it Broadening parameters:} With the  value of metallicity obtained as described above, we  recomputed \vmic, which  is set to minimize the slope of the
relationship between the \FeI\ abundance and the observed EWs. Iteratively, we repeated the analysis with the new set of atmospheric parameters and,  with one $\sigma$ clipping,  we obtained the final values of  \FeI\ and \FeII\ abundances.

\subsubsection{Porto}
\indent {\it Code description:} This method is based on  EWs, which  are measured automatically using ARES\footnote{The ARES code can be downloaded at
http://www.astro.up.pt/}   \citep{2007A&A...469..783S}. These  are then used to compute individual line abundances with MOOG \citep{1973PhDT.......180S}.  The analysis of the Porto node was carried out by S.G. Sousa. 

\indent {\it Iron abundance determination:}
For this exercise we assumed that the excitation and ionization balance is present. In every iteration we rejected outliers above $2 \sigma$. We find the final value of \feh~ when the input \feh\ of the models is equal to the average of the computed line abundances.

\indent {\it Broadening parameters:}
For giants, we computed the microturbulence because it depends on \feh, which is a parameter that we initially set to \feh = 0 for all stars. This was done by determining \feh\ and \vmic\ simultaneously requiring excitation balance. 
For dwarfs, we utilized the value obtained from the GES relation, since it is independent  of the \feh\ of the star. 



\subsubsection{UCM (Universidad Computense de Madrid)}

{\it Code description:} The UCM node relies on EWs.  An automatic code based on some subroutines of 
{\scshape StePar} \citep{tab12} was used to determine   the metallicity. Metallicities are computed using the 2002 version of 
the MOOG code \citep{1973PhDT.......180S}. We modified 
the interpolation code provided with the MARCS grid to produce 
an output model readable by MOOG. 
We also wrote a wrapper program to the MARCS interpolation code to interpolate any
required model on the fly.

{\it Iron abundance determination:} The metallicity is inferred from any previously selected 
 line list. We iterate until the metallicity from
the Fe lines and metallicity of the model are the same.  The EW determination of the Fe lines was carried out with the 
ARES code \citep{2007A&A...469..783S}. 
In addition, we performed a 3-$\sigma$ rejection of the 
$\ion{Fe}{i}$ and $\ion{Fe}{ii}$ lines after a first determination 
of the metallicity. We then re-ran our program again without the rejected lines. This analysis was carried out by J.~I. Gonz\'alez-Hern\'andez, D. Montes, and H.~Tabernero.

{\it Broadening parameters:}  For the van der Waals damping prescription, we use the Uns$\ddot{{\rm o}}$ld 
approximation. 
As in the Porto method, we determined \vmic\ only for giants, while for dwarfs, we fixed \vmic\ by the values obtained from the GES relation.

\begin{table*}[ht!]
\caption{Metallicity of \bs s obtained  individually by each method by analysing neutral iron abundances and assuming LTE. The last two columns indicate the mean value for the microturbulence parameter  obtained by each method, and the standard deviation of this mean. }              
\label{table:BSfeh}      
\centering                                   
\begin{tabular}{c | c c c c c c c c c}          
\hline\hline                        
star  & LUMBA &  Bologna & EPINARBO&  Nice &  UCM &  ULB  &  Porto & \vmic\ (Km/s) & $\sigma$\vmic\\
\hline
\Sco  &  +0.01&  +0.03& --0.10&  +0.00& --0.02& --0.01& --0.02 &1.2& 0.2 \\
\cygA  & --0.42& --0.35& --0.33& --0.25& --0.40& --0.45& --0.39&1.1&0.04\\
\cygB & --0.47& --0.35& --0.48& --0.50& --0.34& --0.74& --0.32&1.1&0.36\\
\alfCenA &  +0.29&  +0.25&  +0.14&  +0.25&  +0.22&  +0.14&  0.23&1.2&0.07\\
\alfCenB &  +0.23&  +0.27&  +0.06&  +0.25&  +0.17&  +0.21&  +0.13&1.1&0.31\\
\alfCet  & --0.13& --0.33& --0.39&  +0.00& --0.38& --0.64& --&1.4&0.4\\
\alfTau  & --0.12& --0.23& --0.31& --0.25& --0.34& --0.43& --&1.4&0.4\\
\Arcturus  & --0.52& --0.56& --0.54& --0.50& --0.50& --0.65& --0.46&1.3&0.12\\
\betAra  &  +0.35&  +0.11& -0.08&  +0.00&  +0.07& --0.16& --&1.5&0.46\\
\betGem  &  +0.05&  +0.07&  +0.03&  +0.00&  +0.16& --0.01&  0.24&1.1&0.21\\
\betHyi & --0.04& --0.06& --0.09& --0.25& --0.11& --0.06& --0.09&1.3&0.04\\
\betVir  &  +0.17&  0.15&  +0.10&  +0.00&  +0.11&  +0.11&  +0.11&1.4&0.09\\
\delEri &  +0.06&  +0.14& --0.06&  +0.00&  +0.04&  +0.00&  +0.00&1.2&0.22\\
\epsEri  & --0.10& --0.11& --0.09& --0.25& --0.15& --0.12& --0.19&1.1&0.05\\
\epsFor & --0.58& --0.59& --0.62& --0.75& --0.68& --0.61& --0.67&1.2&0.13\\
\epsVir  &  +0.09&  +0.09&  +0.02&  +0.00&  +0.24&  +0.04&  +0.08&1.1&0.25\\
\etaBoo  &  +0.34&  +0.30&  +0.33&  +0.00&  +0.08& -0.28&  +0.27&1.4&0.19\\
\gamSge  & --0.01& --0.01& --0.09& --0.25& --0.05& --0.39& -- &1.4&0.34\\
\Gmb & --1.48& --1.47& --1.62& --1.50& --1.48& --1.80& --1.46&1.1&0.57\\
\object{HD107328}  & --0.20& --0.35& --0.26& --0.25& --0.22& --0.47& --0.10&1.2&0.26\\
\object{HD122563} & --2.67& --2.76& --2.76& --3.00& --2.75& --2.84& --2.76&1.3&0.11\\
\object{HD140283}  & --2.51& --2.53& --2.44& --2.50& --2.55& --2.54& --2.57&1.3&0.20\\
\object{HD220009}  & --0.82& --0.77& --0.70& --0.75& --0.79& --0.83& --0.79&1.3&0.14\\
\object{HD22879}  & --0.88& --0.87& --0.91& --1.00& --0.95& --0.83& --0.89&1.2&0.19\\
\object{HD49933}  & --0.43& --0.42& --0.43& --0.50& --0.62& --0.39& --0.49&1.9&0.35\\
\object{HD84937} & --2.22& --2.15& --2.15& --2.00& --2.23& --2.21& --2.21&1.5&0.24\\
\ksiHya & -0.01&  +0.08&  +0.10&  +0.00&  +0.19&  +0.06&  +0.30&1.1&0.32\\
\muAra &  +0.36&  +0.34&  +0.31&  +0.25&  +0.26&  +0.28&  +0.32&1.2&0.13\\
\muCas A  & --0.86& --0.82& --0.82& --1.00& --0.89& --0.78& --0.88&1.1&0.29\\
\muLeo&  +0.37&  +0.39&  +0.31&  +0.25&  +0.50&  +0.23&  +0.34&1.1&0.26\\
\Procyon &  +0.03& --0.03& --0.08&  +0.00& --0.06& --0.01& --0.06&1.8&0.11\\
\psiPhe  & --0.65& --0.57& --0.42&  +0.00& --0.40& --0.47& -- &1.5&0.33\\
\object{Sun}  &  +0.03&  +0.04& --0.06&  +0.00& --0.02& --0.01& --0.03&1.2&0.18\\
\tauCet& --0.51& --0.49& --0.49& --0.75& --0.56& --0.49& --0.56&1.1&0.28\\
\hline
\end{tabular}
\end{table*}

\normalsize

\section{Results}\label{results}

In this section we discuss the metallicity obtained from the three runs described in Sect.~\ref{runs}. This allows us to have a global view of how the different method  compare to each other. We further discuss the impact that our stellar parameters have on the ionization balance, and finally we present the NLTE corrections.

\subsection{Comparison of different methods}
Table~\ref{table:BSfeh} lists the results obtained from \textit{run-nodes}, where every node has determined the metallicity of one spectrum per \bs. The value indicates the result obtained from the analysis of \FeI\ lines under LTE.  The table also lists the mean \vmic\ value obtained by the different nodes, with $\sigma$\vmic\ representing the standard deviation of this mean. In  \fig{fig:run01} we show the difference between the result of each node and the mean literature value as a function of \bs, in increasing order of temperature. The name of the star is indicated at the bottom of the figure, with its corresponding  fundamental temperature at the top of it.

\begin{figure}
  \resizebox{\hsize}{!}{\includegraphics{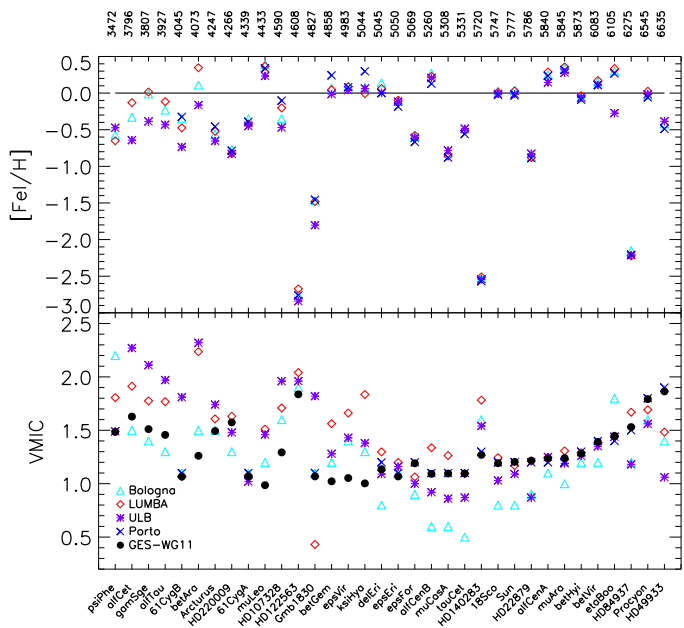}}
  \caption{Metallicity (upper panel) and microturbulence velocity (lower panel) obtained by different methods for each \bs, as a function of temperature. Black dots correspond to the values of \vmic\ obtained from the GES relation of Bergemann and Hill.}
  \label{fig:run01_vmic}
\end{figure}

For warm stars (i.e. \teff $> 5000$~K) the values of  metallicity obtained by the different methods have a  standard deviation of 0.07~dex. Moreover, these values agree  well with the  literature, with a  mean offset of +0.04~dex. The standard deviation increases notably for cooler stars, being typically on the order of 0.1~dex, with a maximum of 0.45 for \betAra. Note that this star has a literature value that was determined from  photographic plates \citep{1979ApJ...232..797L} and is thus uncertain. A similar behavior can be seen in \fig{fig:feh_pastel} with the values reported in the literature, where \feh\ of cold stars  present more scatter than hot stars.  The fact that obtaining a good agreement in \feh\ for cool stars is more difficult than for warm stars is mainly due to line crowding and the presence of molecules in the spectra of very cool stars. This means that the iron lines in most of the cases are not well recognized  nor well modeled. Moreover, absorption lines in cold stars can be very strong, making the continuum normalization procedure extremely challenging.  Also, 3D effects can become important  in  giants \citep[e.g.][]{2007A&A...469..687C, 2010A&A...515A..12C} and our models consider only 1D.

Note that for some stars, like \betAra, 61~Cyg~A and B, \Gmb\ and \object{HD122563}, we obtain a fair agreement in metallicity.  The mean value, however, differs significantly from the  mean literature value. In Sect.\ref{lit} we discussed how the \feh\ from the different works can differ significantly due to inhomogeneities between the different works. A more detailed discussion of each star, especially those with significant discrepancies compared to the mean literature value, can be found in Sect.~\ref{discussion}.

When using 1D static models to determine parameters we need to employ  additional broadening parameters (micro- and macroturbulence velocity), which represent the non-thermal motions in the photosphere. Since these motions are not described in 1D static atmosphere models, broadening parameters become important to compensate for the effects of these motions.  Figure~\ref{fig:run01_vmic} shows the correlation between \feh\ and \vmic\ for the Bologna, LUMBA, ULB and Porto methods.   \citet{1981A&A....97..145N} made an analysis of \vmic~ as a function of \feh, \teff\ and \logg\ for solar-type dwarfs obtaining a relation where \vmic\ increases as a function of \teff, which agrees with our results of \vmic~ shown in \fig{fig:run01_vmic} for warm stars (\teff $ \geq 5000$ K).  This effect has also been noticed in \citet{2005AJ....129.1063L} and \citet{2012MNRAS.423..122B}. Metal-poor stars  are outliers of the smooth relation, with \object{HD140283} being the most evident one. Such metal-poor stars were not included in the samples of \citet{1981A&A....97..145N} and \citet{2012MNRAS.423..122B}. The microturbulence velocity decreases as function of \teff\ for stars cooler  than \teff $\sim$ 5000 K, although with a larger scatter than for warm stars. This general behavior agrees with the GES relation (see \sect{nodes}), which is plotted with black dots in \fig{fig:run01_vmic}.

Note that although each method shows the same behavior of \vmic\ as a function of temperature, the absolute value of \vmic\ differs. The differences found between methods in \vmic\ help to achieve a better general agreement of \feh. 

\begin{figure}
  \resizebox{\hsize}{!}{\includegraphics{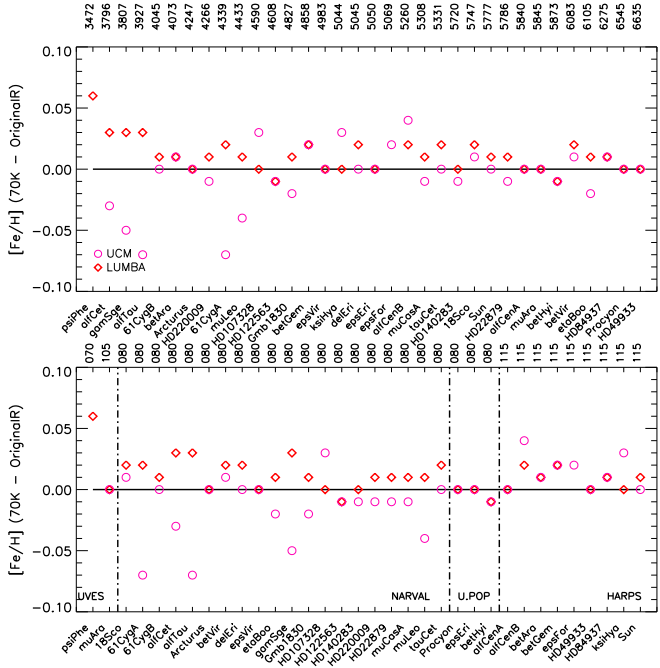}}
  \caption{Difference of metallicity obtained from 70~k and Original library for UCM and LUMBA methods.  Upper panel: difference as a function of \bs\ temperature. Lower panel: difference for stars of same instrument. }
  \label{fig:run02}
\end{figure}

\subsection{ Comparison of different resolutions}
In \fig{fig:run02} we plot the comparison of the results from LUMBA and UCM obtained for \feh\ when considering the 70~k and original library. 

As in previous figures, we illustrate the difference in metallicity as a function of \bs,  in order of  increasing temperature in the upper panel. 
In the lower panel of \fig{fig:run02} we plotted   together the stars observed with the same instrument. Different instruments are separated by the dashed line. The value of the spectral resolution before convolution is indicated at the top of the figure.

It is interesting to comment on the result of  \psiPhe, which has the lowest original resolution and is the coldest star, because it shows the greatest difference. In the case of the LUMBA method, the synthetic spectra produced by SME need to have a given resolving power, which is set to be constant along the entire spectral range. This is, in the original spectra, not completely true. In this particular case,  the upper part of the CCD  of the UVES spectrum has a  resolution that is lower than 70000 (see Paper~II). In any case, the difference is of about 0.06~dex, which is negligible compared to the uncertainty obtained for this star of about 0.5 dex (see \tab{table:BSfeh_final} and \sect{errors}).

The same can happen for the results from the original NARVAL spectra, which we assume to be $R = 80000$. As discussed in Paper~II, the resolving power of NARVAL might not be exactly $80000$, but it is acceptable to assume  initially a constant resolving power of $R=80000$ for all the original spectra for creating the 70~k library.  However, when  analyzing  directly the original spectra with SME, wavelength-dependent deviations from the constant input resolution might affect might affect the results, explaining the scatter around the zero line observed in \fig{fig:run03} for NARVAL spectra. A discussion of the impact of parameters when the exact resolution of spectra is not given can also be found in \citet{2011A&A...525A..71W}. UVES-POP spectra, on the other hand, have a well defined resolving power and our results agree very well. Finally, HARPS spectra also have a quite well established original resolution. It is also the highest resolution of our sample. 

It is worth to comment on the results obtained by UCM for cool stars, where the difference between the original and convolved spectra are larger than for warm stars. This effect can be attributed to the  contribution of lines other than Fe that can be better resolved at higher resolution, producing a slightly different measurement of the EW. In general, differences of less than 0.03~dex are present for both methods when using different resolutions (and SNR), which is within the errors obtained in the abundances (see \sect{errors}).

\subsection{Comparison of different instruments}
\begin{figure}
  \vspace{-1.8cm}
  \resizebox{\hsize}{!}{\includegraphics{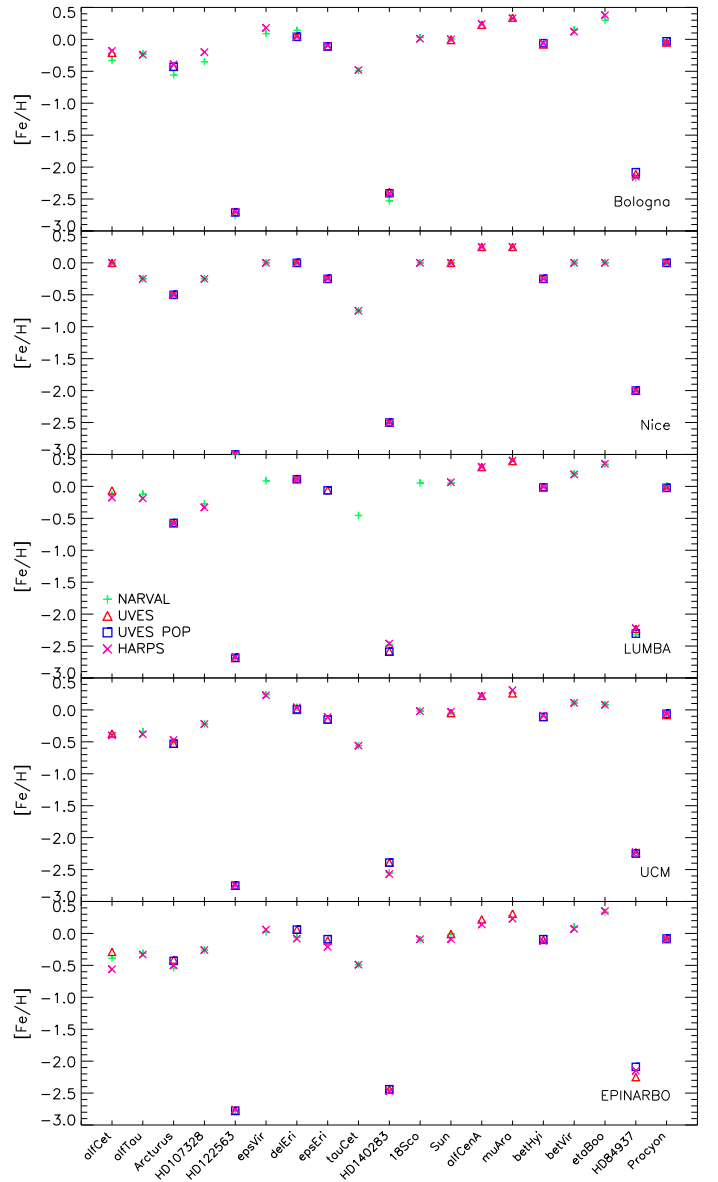}}
  \vspace{-1.5cm}
  \caption{Metallicity of \bs~as a function of effective temperature. Symbols represent different instruments (see legend). Each panel shows the result of one method, indicated in each panel.}

  \label{fig:run03}
\end{figure}

For many of the \bs s, we have more than one observation. We expect our results to be consistent under different instruments. For that reason, we determined \feh\ for each spectrum in the complete 70~k library separately and compared them. The results obtained for the methods of Nice, Bologna, EPINARBO, UCM, and LUMBA are displayed in \fig{fig:run03}.  The figures present the value of the metallicity as a function of \bs, with increasing temperature.

There is a general good agreement when different spectra are analyzed for the same star. \Procyon, which has observations in every instrument from our library, has an excellent agreement for each method considered here. 
On general,  our results and data are consistent because we do not find signature of one particular instrument giving systematic differences. In the same way, we do not find the result of one particular star being biased towards one observation.  This comparison shows also that the data reduction software of the spectrographs perform correctly.

\subsection{Self consistency and ionization balance}\label{ionization_balance}

Usually, when determining parameters, \teff, \logg,
\vmic~(and \vmac~in case of synthetic spectra) and [Fe/H]  must be chosen such that the iron
abundance obtained from neutral lines agrees with that obtained from ionized
lines, the so-called ionization balance.  Corresponding constraints are used to find the
best \teff~ (a flat trend of \FeI~ with excitation potential) and \vmic~ (a flat
trend of \FeI~ with EW).

Since in this particular work we do not change \teff\ and \logg, the simultaneous determination of the other
parameters becomes  the dominant means for approaching ionization and excitation balance. For methods based on EWs, \vmic~ helps
to obtain abundances in a line-to-line approach that doe not depend on the
reduced EW or wavelength range. For methods based on synthetic
spectra, \vmic\ and \vmac\ are treated as broadening parameters that help to improve the fit
of the synthesis to observed line profiles. 

\begin{figure}
  \resizebox{\hsize}{!}{\includegraphics{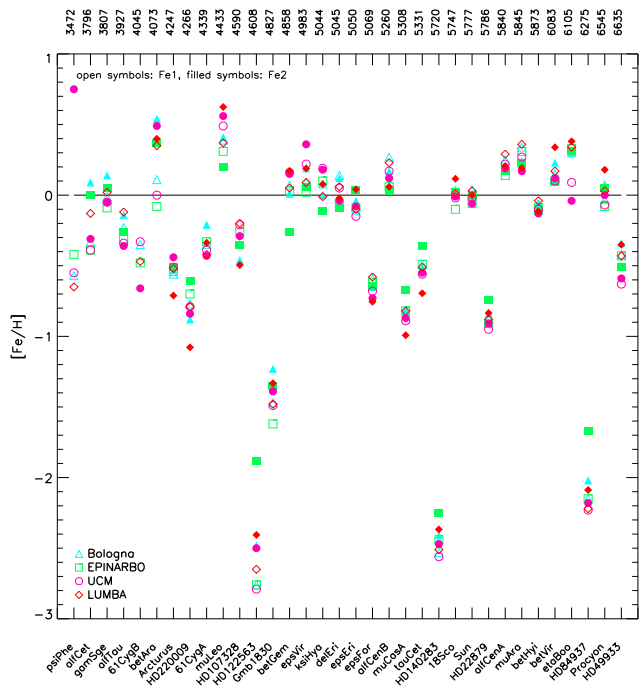}}
 \vspace{0.2cm}
  \caption{Neutral and ionized iron abundances obtained for \bs s as a function
  of effective temperature by different methods (see legend). Open symbols
  represent \FeI\ abundances while filled symbols represent \FeII\ abundances. }
  \label{fig:fe12}
\end{figure}

Since \teff\ and \logg\ are taken from fundamental relations and are independent of
spectral modeling, ionization balance and the mentioned relations tell us how
well our models are able to reproduce our observations. Figure~\ref{fig:fe12}
displays the iron content obtained from neutral and ionized lines for the \bs s
using EPINARBO, UCM,  Bologna and LUMBA methods. The stars have been plotted with
increasing temperature and each symbol represents one method.  Open and filled symbols
indicate \FeI\ and \FeII\ abundances, respectively. 

Generally, all nodes  show a significant difference between \FeI\ and \FeII\ abundances for
\object{HD122563}, \Gmb\ and \muAra. For other cases, such as \betGem, only some methods show large differences while others show an agreement. Cool stars like \alfTau\ or \alfCet\ are also problematic because the available 
\FeII\ lines are often blended by molecules and it becomes difficult to
model them with our current theoretical input data. In fact, it was impossible to
create a \FeII\ line mask for \psiPhe\ when analyzed with the LUMBA method.  The \FeII\ abundances obtained for the coolest stars by any method can thus be unreliable. To be able to obtain reliable \FeII\ abundances for such stars, the synthesis methods would need to have a list of molecules capable of reproducing those blends. 

\begin{figure}
  \resizebox{\hsize}{!}{\includegraphics[angle=270]{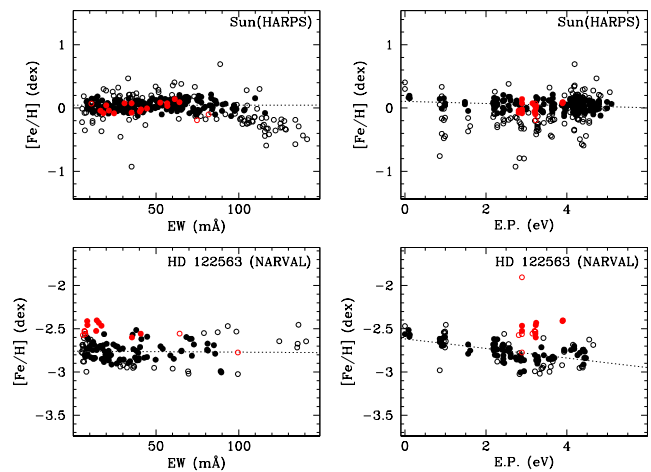}}
    \caption{GALA outputs of the Bologna method for the \object{Sun} (HARPS, upper panels)
     and \object{HD122563} (NARVAL, lower panels) for the {\em run-nodes} test. In all
     panels, black symbols refer to \FeI\ and red ones to \FeII, while empty
     symbols refer to rejected lines (see Sect.\ref{bol}) and solid ones to lines effectively used for the analysis. A
     dotted line shows the result of a linear fit to the used \FeI\ lines in all
     panels.}
  \label{fig:fe12_gala}
\end{figure}

Figure~\ref{fig:fe12_gala} shows the trends of the iron abundance as a function of EW
and excitation potential for the \object{Sun} (a good case) and \object{HD122563} (an unbalanced
case) as obtained by the Bologna node (see also Sect.~\ref{bol}). Black and red dots correspond to neutral and ionized iron abundances, respectively. The figure
 shows that a perceptible difference between \FeI\ and \FeII\ abundances
results when using \logg\ from Table~\ref{table:BMKparams}, and also a trend of
iron abundance with excitation potential appears when using the \teff\ from the
same table. If the parameters were let free, as in the traditional EW-based
method, both gravity and temperature would have to be re-adjusted to obtain
self-consistent results.

Even in the good cases, where the abundances of neutral and ionized iron are well
determined, a small difference between the two can appear and it is often
difficult to reconcile \FeI\ and \FeII\ abundances. \citet{2011ApJ...743..135R},
in their attempt to review the fundamental parameters of \Arcturus\ with a method
very similar to the one presented in this work, obtained a difference of 0.12~dex between \FeI\ and \FeII\
abundances,  which is explained as a limitation of the 1D-LTE models, that cannot reproduce
the data well enough. Similarly, \citet{2003AJ....125.2085S}  reported
problems in their analysis of the open cluster M~34, where \teff~and \logg~ were
kept fixed to values obtained from the color-magnitude diagram and the final
iron abundance from ionized and neutral Fe lines did not fully satisfy ionization
balance, especially in the case of the coldest K dwarfs. An extensive discussion
on this subject  can be found in \citet{2004A&A...420..183A}, who analyzed field
stars in the solar neighborhood. Their Figure~8 shows the differences obtained
from neutral and ionized lines of iron and calcium, where differences can reach
0.5~dex in the most metal-rich cases. They argue that, to satisfy ionization
balance, dramatic modifications of the stellar parameters are necessary, which
would be translated to unphysical values. All
aforementioned works explain this effect as due to departures from LTE, surface
granulations, incomplete opacities, chromospheric and magnetic activity, and so
on.  For an extensive discussion on this issue for five of our \bs s (the \object{Sun}, \Procyon, \object{HD122563}, \object{HD140283}, \object{HD84937} and \object{HD122563}) see also \citet{2012MNRAS.427...27B}.

\begin{figure}
  \resizebox{\hsize}{!}{\includegraphics{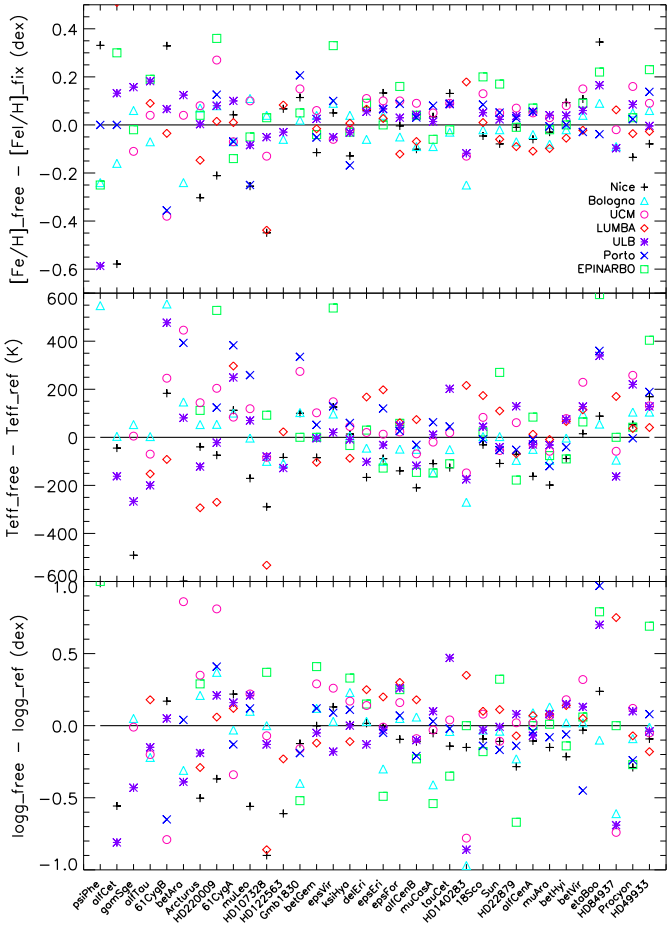}}
  \caption{Difference in metallicity (upper panel), effective temperature (middle
  panel) and surface gravity (lower panel) of \bs s as obtained by different
  methods, between free and fixed analysis (see text).}
  \label{fig:free}
\end{figure}

We performed an additional abundance analysis determining simultaneously \teff\ and \logg, together with \feh\ and \vmic\ on
the 70~k~library.  
Our idea was to quantify by how much must \teff~ and \logg~ be altered in order to
obtain excitation and ionization balance in each method. The results of this "free" analysis are illustrated in \fig{fig:free}, where the difference between the
"fixed" (determination of \feh\ via fixing \teff\ and \logg) and the "free"
analysis are shown for each \bs. Metallicity, temperature and surface gravity are
plotted in the upper, middle and lower panel of \fig{fig:free}, respectively.

As expected, the metallicity obtained when forcing ionization equilibrium for 1D LTE models
is different from that obtained with the fundamental \teff\ and \logg. The median difference in metallicity for solar-type stars  is smaller than for 
the coldest, hottest and metal-poor stars. 
The differences obtained are usually related  to larger deviations in \teff\ and \logg\
from the fundamental value, as seen in \fig{fig:free} and also discussed in e.g. \citet{2004A&A...420..183A}
and \citet{2011ApJ...743..135R}. In \Gmb, for example, the results of
\teff\ and \logg\ from the free spectral analysis agree better with what has been
reported in  PASTEL \citep{2010A&A...515A.111S}, which is more
than 250 K above the fundamental value. \object{HD140283} is another case where the free
temperature and surface gravity are 200~K and 0.7~dex smaller than the
fundamental value, resulting in a \feh\ that is $\sim 0.2$ dex more metal-poor
than the fixed case. On the other hand, the smallest differences in \feh\ are
related to small deviations in \teff\ and \logg. Examples of this cases are
\muCas~A, \alfCenA, \alfCenB and the \object{Sun}. 

In general, when looking at the results of individual methods, a difference of up to 200~K in \teff\ and 0.25~dex in \logg\ would
be necessary to restore excitation and ionization balance in the problematic \bs s. This would
introduce a change of $\sim 0.1$ dex in metallicity as well.  It is important to comment that this test is just an illustration of the effects of freeing \teff\ and \logg\ to retrieve ionization balance but does not represent the real performance of the different methods when determining three parameters.  Here we are only concentrating in the analysis of iron lines and not the analysis of other important spectral features that can affect the determination of \teff\ and \logg. This can have important consequences for methods based on  SME or  MATISSE, for example.  A full explanation of the performance of the methods in the parametrization of UVES spectra will be found in Smiljanic et al. (in prep).

\subsection{NLTE corrections}\label{nlte}

Recently, \citet{2012MNRAS.427...27B} presented a thorough investigation of the \FeI-\FeII\ ionization balance in five of the Benchmark Stars included here (\object{Sun}, \Procyon, \object{HD122563}, \object{HD84937}, \object{HD140283}) and one more extremely metal-poor star (G64-12). In particular, they utilized an extensive Fe model atom and both traditional 1D and spatially and temporally averaged 3D hydrodynamical models to assess the magnitude of NLTE effects on Fe line formation. \citet{2012MNRAS.427...27B} concluded that only very minor NLTE effects are needed to establish ionization balance at solar metallicities, while very metal-poor stars imply effects on the order of +0.1 dex on \FeI\ lines. \FeII\ lines are everywhere well modelled by the LTE assumption.
  \begin{figure}
 \resizebox{\hsize}{!}{\includegraphics{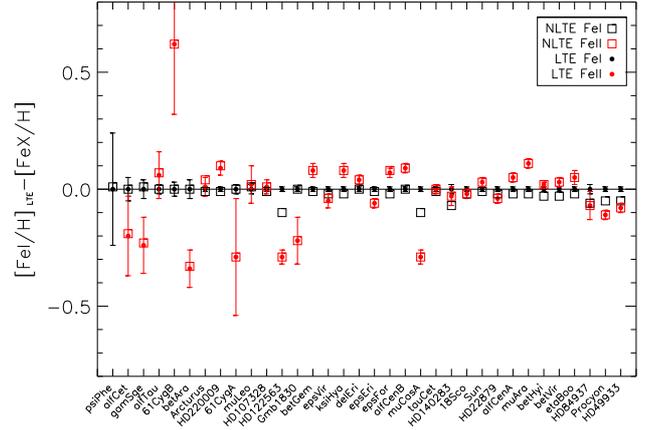}}
  \caption{Difference of final [\FeI /H](black) and [\FeII /H] (red) for each \bs. Squares show the abundances after NLTE corrections.  Error bars represent the uncertainties coming from the line-to-line scatter and the uncertainties coming from the associated uncertainties in \teff, \logg\ and \vmic\  (\sect{errors})}
  \label{fig:NTLE-LTE}
\end{figure}

The NLTE calculations were extended by \citet{2012MNRAS.427...50L} to cover a large cool star parameter space. Here, we interpolated within the grid of NLTE corrections by \citet{2012MNRAS.427...50L}  the stellar parameters adopted for each Benchmark Star as taken from \tab{table:BMKparams}. Each Fe line used in the final \feh\ determination was corrected individually. When a NLTE correction was not available for a specific line, we used the median of the corrections computed for all other lines.  This is possible to do as the corrections for all lines of a particular star are very similar, as shown by \citet{2012MNRAS.427...27B}. The difference between the final  Fe abundances for single and ionized lines is visualized in \fig{fig:NTLE-LTE} for each star (see \sect{errors} for details of how the final abundances are determined). The stars are plotted in  order of increasing effective temperature. Black  indicates that the iron abundance is determined from \FeI\ lines while red indicates that the abundance is determined from \FeII\ lines.  Dots and square symbols indicate the LTE and NLTE abundances, respectively. The errors bars  are plotted only for the LTE abundances, as they do not change after NLTE corrections.  The errors considered in this plot correspond to the sum of the scatter found for the line-by-line abundance determination and the errors obtained considering the associated uncertainties in the fundamental parameters (see \sect{errors} for details).

In general, NLTE corrections can vary between -0.10 to +0.15 dex for individual lines, but on average the departures of NLTE affect the metallicity by <0.05 dex for all stars. Exceptions are the hottest stars and the most metal-poor ones, which can differ up to 0.1 dex.  Since the corrections due to NLTE effects are small, even when looking at the final NLTE abundances in \fig{fig:NTLE-LTE}, we still find cases where ionization imbalance is significant, especially for the cold stars. We conclude that neglecting NLTE effects is not a likely explanation for the ionization imbalance.

\section{The metallicity determination}\label{errors}

Since each method and corresponding criterium used to give a final \feh\ value  differ, we combine our results by looking at individual abundances in a line-by-line approach. Since the Nice method is based on a global fitting of a whole section of the spectrum, abundances of individual lines for that method are not provided.  
We note that the  setup employed by the LUMBA node for this analysis performed a simultaneous fit of all pixels contained in the specified line mask, and thus it did not provide abundances of individual lines per se. However,  LUMBA employed a post-processing code, that  determined best-fit $\log gf$ values for each line. This is equivalent to determining best-fit abundances. The resulting $\log gf$ deviation from the nominal value is then added to the global metallicity of each star derived by SME in order to reconstruct individual line abundances.

We performed several steps to combine and thus determine the metallicity of each star. This analysis was mostly carried out by P. Jofr\'e, U. Heiter, J. Sobeck and K. Lind. 

Firstly, we selected those lines with $\log{(\mathrm{EW}/\lambda)} \leq -4.8$. The objective was to use lines which are on the linear part of the curve of growth, in order to  avoid saturated lines and mitigate the effect of ``wrong microturbulence" and ``wrong damping parameters'' which affect strong lines.  
The transition from the linear part to the saturated part of the curve of growth occur at $\log(\mathrm{EW}/\lambda) \sim -5.0$, more or less independent of stellar parameters \citep[See e.g. Figs. 16.1 to 16.6 of ][or \citealt{1987Ap&SS.136..351V}]{2005oasp.book.....G}. The transition point is slightly above for cool models, while  slightly below  for hot models.  In addition, the transition value was checked for each Benchmark Star by constructing empirical curves of growth from the output of the Bologna method. For the different kind of stars presented here, the limit of -4.8 seems to be a good compromise between the number of lines and the saturation criterion.

Secondly, we calculated the mean and standard deviation of  all abundances and selected those lines that were analyzed by at least three different groups and for which the values agreed within 2 $\sigma$ with the mean abundance. 

Thirdly, we calculated the mean abundance from the different methods for each selected line. For consistency checks on metallicities, each abundance was plotted as a function of wavelength, EW and excitation potential (E.P.) to account for excitation balance. The relations can be found in Fig.~\ref{fig:group1}, \ref{fig:group2}, \ref{fig:group3}, \ref{fig:group4} and \ref{fig:group5}.  Additionally, NLTE corrections were  applied individually for each selected line and star (see \sect{nlte}).  An extensive discussion is found in \sect{discussion}

Finally, we computed the final value of \FeI\ and \FeII\ abundances from the average of the selected lines. To compute the final metallicity, we considered the value of 7.45 for the absolute solar iron abundance from \citet{2007SSRv..130..105G}.  The final value of \feh\ obtained from \FeI\ lines after corrections by NLTE effects is listed in the second column of \tab{table:BSfeh_final}. The third column indicates the standard deviation of the abundances obtained from the selected \FeI\ lines.  The list of lines selected for each star can be found as part of the online material.

\subsection{Errors due to uncertainties in \teff, \logg\ and \vmic}

We are basing our analysis on fixed values for \teff\ and \logg, but these values have associated errors that give the metallicity an additional uncertainty. In a similar manner, we want to study the effect on the final metallicity due to the uncertainties in the \vmic\ parameter. To quantify the error of \feh\ due to the associated errors in \teff, \logg\ and \vmic, we performed additional runs determining the iron abundances using the same setup as described for {\it run-nodes} in \sect{method}, but changing the input value of \teff, \logg\ and \vmic\ by considering  $\mathrm{T_{eff} \pm \Delta \mathrm{T_{eff}}}$, $\log g \pm \Delta \log g$ and $v_{\mathrm{mic}} \pm \Delta v_{\mathrm{mic}}$, respectively. The values of  $\Delta \mathrm{T_{eff}}$ and $\Delta \log g$ can be found in \tab{table:BMKparams} and were determined in Paper~I, while for the value of $\Delta v_{\mathrm{mic}}$ we considered the scatter found by the different nodes from the standard {\it run-nodes} which can be found in the last column of \tab{table:BSfeh}. 

This analysis gave us 6 additional runs, which were performed by the methods LUMBA, EPINARBO, Porto, UBL and UCM. To be consistent with our main results,  we determined the iron abundance of only the lines that passed the selection criteria after the main run. The final differences of $(\mathrm{\feh}_{\Delta^{-}}-\mathrm{\feh}_{\Delta^{+}})$, where $\mathrm{\feh}_{\Delta^{\pm}}$ correspond to the metallicities obtained considering the parameters $\pm$ their errors, for \teff, \logg\ and \vmic\, respectively. These values are also listed in \tab{table:BSfeh_final} for each star.

\subsection{Discussion}\label{discussion}

To understand better our results,  we divided the stars into 5 groups: metal-poor stars, FG dwarfs, FGK giants, M giants, and K dwarfs. Each group is discussed separately in the following sections.

\subsubsection{Metal-poor stars}\label{metal-poor}
This group includes the stars \object{HD122563}, \object{HD140283} and \object{HD84937}. Our results agree well with an internal scatter in a line-by-line approach of about 0.12 dex before the  line selection process described in \sect{errors}.  
A similar differential analysis between the results obtained for atmospheric parameters from equivalent widths and synthetic spectra on high resolution spectra of metal-poor stars was done by \citet{2010A&A...517A..57J}. In that study, 35 turn-off metal-poor stars were analyzed using the same data and line list and different atmosphere models. The general scatter was 0.13 dex in metallicity when \logg~and \teff~were forced to agree by 0.1~dex and 100~K, respectively.  Although here we determine only metallicity, it is encouraging to obtain a mean scatter of 0.06~dex when considering the independent results of the seven methods.

\begin{figure}
\resizebox{\hsize}{!}{\includegraphics[angle=90]{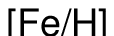}}
  \caption{Trends of abundances as a function of excitation potential (left panels) and reduced equivalent width (right panels) in the group of metal-poor stars.}
  \label{fig:group1}
\end{figure}

The abundances of the selected lines for each metal-poor star as a function of E.P. are shown in the left panels of \fig{fig:group1}, while the abundances as a function of reduced EW are shown in the right panels of the figure. Black dots correspond to \FeI\ abundances, corrected by NLTE effects as described in \sect{nlte}, while the red dots correspond to the \FeII\ abundances. The solid red and black horizontal lines indicate the averaged \FeII\ and \FeI\ abundance, respectively.  In addition, we plotted with a dot-dashed line the linear regression fit to the \FeI\ abundances, where its slope and  error are written in the bottom of each panel.

In metal-poor stars the continuum is easy to identify, although other difficulties appear, such as the low number of iron lines detectable in the spectra, especially those of ionized iron. In our case, the common lines that passed the selection criteria explained above can be seen in \fig{fig:group1}.  \object{HD84937} is the most extreme case, where we have only 1 ionized and 20 neutral iron lines that are used for the final \feh\ determination. 

NLTE effects can change significantly the metallicity of metal-poor stars \citep{1999ApJ...521..753T, 2005ARA&A..43..481A}.  After applying NTLE corrections to our selected LTE \FeI\ abundances, the metallicities increase by up to approximately 0.1~dex, which agree with the  investigation of  \citet{2012MNRAS.427...27B}  for these three Benchmark Stars.

The largest difference between \FeI\ and \FeII\ abundances is for the metal-poor giant \object{HD122563}. However, in the distribution of \FeI\ lines as a function of E.P. one can see a significant slope in the regression fit of $- 0.066\pm 0.008$.  The regression fit as a function of EW shows a slope of 0.07 that can be neglected when considering the error of 0.11. Since those fits are obtained after making the NLTE corrections, we attribute this trend to 3D effects, which are most important for cool metal-poor stars \citep[e.g.]{1999A&A...346L..17A, 2007A&A...469..687C}. See also \citet{2012MNRAS.427...27B} for the study in this regard of \object{HD122563}.   The second metal-poor star, \object{HD140283} also presents  a negative slope for \FeI\ abundances as a function of E.P., although it is less pronounced and its error is larger than the case of \object{HD122563}. It is interesting to see that for this metal-poor subgiant we obtain a good ionization balance.  The last metal-poor star of our group, \object{HD84937}, presents  quite a flat regression fit when looking at the abundances as a function of E.P or EW and considering the errors. Moreover, \FeI\ and \FeII\ abundances agree when the errors due to \teff\ and \logg\ are taken into account.  

We conclude that although one should be aware that there is a large ionization and excitation imbalance for \object{HD122563}, we can average the abundances and obtain robust values of metallicities for metal-poor stars given their fundamental parameters and associated errors.

\begin{table*}[ht!]
\caption{Final metallicity of \bs s obtained via combination of individual line abundances of neutral lines corrected by NLTE effects. The metallicity is associated with different sources or errors: standard deviation of the line-by-line abundance of the selected \FeI\ lines ($\sigma$ \FeI);  errors due to the uncertainty in \teff, \logg\ and \vmic, ($\Delta$ (\teff), $\Delta$ (\logg), $\Delta$ (\vmic), respectively). Error due to difference between NLTE and LTE \FeI\ abundance ($\Delta$ (LTE) ); error due to difference between \FeI\ and \FeII\ abundance $\Delta$ (ion); and  standard deviation of the line-by-line mean  of \FeII\ abundance ($\sigma$ \FeII).  The last two columns indicate the number of selected lines used for the determination of \FeI\ and \FeII\ abundances, respectively.}
\label{table:BSfeh_final}      
\centering      
\begin{tabular}{c | c c c c c c c c c c}         
\hline\hline                        
star & \feh &  $\sigma$ \FeI\ & $\Delta$ (\teff)& $\Delta$ (\logg) & $\Delta$ (\vmic) & $\Delta$ (LTE) & $\Delta$ (ion) & $\sigma$ \FeII & N \FeI\ & N \FeII \\
\hline
{\bf Metal-Poor} \\
\object{HD122563}&  --2.64& 0.01& 0.02& 0.00& 0.01&   +0.10&  --0.19& 0.03& 60 & 4\\
\object{HD140283}&  --2.36& 0.02& 0.04& 0.02& 0.00&   +0.07&   +0.04& 0.04& 23& 2\\
\object{HD84937}&  --2.03& 0.02& 0.04& 0.02& 0.01&   +0.06&  --0.01& -- &20  & 1\\
\hline
{\bf FG dwarfs}\\
\delEri&   +0.06& 0.01& 0.00& 0.00& 0.01&   +0.00&   +0.04& 0.02& 156 &11\\
\epsFor&  --0.60& 0.01& 0.01& 0.00& 0.00&   +0.02&   +0.09& 0.02&148 & 8\\
\alfCenB&   +0.22& 0.01& 0.01& 0.00& 0.02&   +0.00&   +0.09& 0.02&147 &9\\
\muCas&  --0.81& 0.01& 0.01& 0.01& 0.01&   +0.01&   +0.01& 0.02&145 &7\\
\tauCet&  --0.49& 0.01& 0.00& 0.00& 0.00&   +0.01&   +0.01& 0.02&148 &10\\
\Sco&   +0.03& 0.01& 0.01& 0.00& 0.01&   +0.02&   +0.00& 0.02&158 &10\\
Sun&   +0.03& 0.01& 0.00& 0.00& 0.00&   +0.01&   +0.04& 0.02&150 &9\\
\object{HD22879}&  --0.86& 0.01& 0.03& 0.01& 0.01&   +0.02&  --0.02& 0.02& 117&10\\
\alfCenA&   +0.26& 0.01& 0.01& 0.00& 0.00&   +0.02&   +0.07& 0.02& 150&12\\
\muAra&   +0.35& 0.01& 0.00& 0.00& 0.00&   +0.02&   +0.13& 0.02&143 & 13\\
\betHyi&  --0.04& 0.01& 0.01& 0.00& 0.00&   +0.03&   +0.05& 0.01&143 &12\\
\betVir&   +0.24& 0.01& 0.01& 0.00& 0.01&   +0.03&   +0.06& 0.02& 148& 10\\
\etaBoo&   +0.32& 0.01& 0.00& 0.00& 0.01&   +0.02&   +0.07& 0.03& 127 &10 \\
\Procyon&   +0.01& 0.01& 0.01& 0.00& 0.00&   +0.05&  --0.06& 0.02&135&12\\
\object{HD49933}&  --0.41& 0.01& 0.04& 0.02& 0.02&   +0.05&  --0.03& 0.02& 93 &6\\
\hline
{\bf FGK giants}\\
\Arcturus&  --0.52& 0.01& 0.00& 0.00& 0.06&   +0.01&   +0.02& 0.04&151 &10\\
\object{HD220009}&  --0.74& 0.01& 0.01& 0.00& 0.07&   +0.01&   +0.10& 0.03& 148&11\\
\muLeo&   +0.25& 0.02& 0.00& 0.00& 0.13&  --0.01&   +0.01& 0.08& 139&11\\
\object{HD107328}&  --0.33& 0.01& 0.01& 0.00& 0.16&   +0.01&   +0.02& 0.03&137 &11\\\
\betGem&   +0.13& 0.01& 0.01& 0.00& 0.13&   +0.01&   +0.09& 0.03& 146&13\\
\epsVir&   +0.15& 0.01& 0.02& 0.00& 0.15&   +0.02&  --0.03& 0.03&139 &12\\
\ksiHya&   +0.16& 0.01& 0.01& 0.00& 0.17&   +0.02&   +0.10& 0.03&151 &11\\
\hline
{\bf M giants}\\
\psiPhe&  --1.24& 0.24& 0.05& 0.03& 0.30&  --0.01&  -- &  -- &  23&0\\
\alfCet&  --0.45& 0.05& 0.17& 0.08& 0.34&   +0.00&  --0.20& 0.17&35 &3\\
\gamSge& --0.17& 0.04& 0.13& 0.09& 0.22&  --0.01& --0.25& 0.12&29&4\\\
\alfTau&  --0.37& 0.02& 0.02& 0.02& 0.12&   +0.00&   +0.06& 0.10&76&9\\
\betAra&  --0.05& 0.04& 0.01& 0.04& 0.16&   +0.00&  --0.34& 0.08 &62&8\\
\hline
{\bf K dwarfs}\\
\cygB&  --0.38& 0.03& 0.01& 0.01& 0.01&   +0.00&   -- & -- &119& 2\\
\cygA&  --0.33& 0.02& 0.00& 0.01& 0.00&   +0.00&  --0.29& 0.25&138&3\\
\Gmb&  --1.46& 0.01& 0.05& 0.03& 0.30&   +0.00&  --0.22& 0.10&116&4\\
\epsEri&  --0.09& 0.01& 0.00& 0.00& 0.00&   +0.01&  --0.05& 0.02&153&11\\

\hline \hline
\end{tabular}
\end{table*}

  \begin{figure}
 \resizebox{\hsize}{!}{\includegraphics[angle=90]{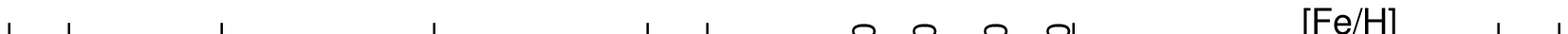}}
  \caption{Trends for group of FG dwarfs. } 
  \label{fig:group2}
\end{figure}

\subsubsection{FG dwarfs}\label{FG}
The stars \delEri, \epsFor, \alfCenA\ \& B, \muCas, \tauCet, \Sco, \object{Sun}, \object{HD22879}, \muAra, \betHyi, \betVir, \etaBoo, \Procyon\ and \object{HD49933} belong to this group.    The mean internal $1 \sigma$ scatter of these stars when looking at all abundances of individual lines is of 0.13~dex, while the value when looking at the results of the individual methods is of 0.07~dex. Moreover, our results agree within 0.04~dex with the literature, as seen in \fig{fig:run01}.  Note that the final line-to-line scatter for these stars is reduced to $\sim 0.01$ from the initial scatter after our selection of lines. NLTE corrections for these stars are very small, usually less than 0.03~dex, with the exception of \Procyon\ and \object{HD49933}, which are of the order of 0.05~dex (see \fig{fig:NTLE-LTE}). These stars have high effective temperatures, which produce greater departures from LTE than cool stars \citep{2012MNRAS.427...27B}. 

As in the case of the metal-poor group, we have plotted the abundances of the selected lines for each star as a function of E.P. and reduced EW in \fig{fig:group2}. This group shows that our selected lines are well-behaved, in the sense that excitation and ionization balance are in general satisfied. Usually a difference between ionized and neutral iron abundances is less than 0.1~dex for this group of stars, which can be confirmed with \fig{fig:NTLE-LTE}. There are few exceptions, such as the hot star \Procyon, and the solar-type stars \epsFor, \alfCenB\ and  \muAra. The latter presents the larger ionization imbalance, which can be explained by the rather large excitation imbalance (with a slope of $-0.012 \pm 0.008$~dex in the regression fit as a function of E.P.). We find no significant trend as a function of $\log(\mathrm{EW}/\lambda)$ when considering the errors of the regression fits. Note that the hot stars \Procyon\ and \object{HD49933} also present a significant excitation imbalance in the regression fits. 

Recently, \cite{2012ApJ...757..161T} made a comparative spectral analysis of FG dwarfs using three different methods to determine parameters. Two of their methods overlap with our own, namely SME (LUMBA) and MOOG (UCM, Porto and EPINARBO). They obtained a systematic difference of $0.068 \pm 0.014$~dex in metallicity when analyzing 31 stars with these two methods, which is attributed to the different \teff\ and \logg\ obtained from the simultaneous analysis, the different way of placing the continuum, and the different lines used by each methods. 

We conclude that it is acceptable to average the abundances of our selected lines and that we are able to provide robust results for \feh\ for FG dwarfs based on their fundamental temperature and surface gravity.

  \begin{figure}
\resizebox{\hsize}{!}{\includegraphics[angle=90]{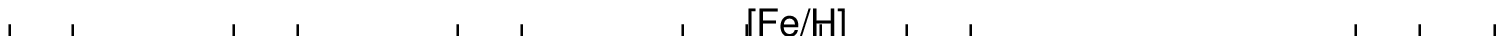}}
  \caption{Trends for group of FGK giants. }
  \label{fig:group3}
\end{figure}

\subsubsection{FGK giants}\label{RG}
These are \Arcturus, \muLeo, \betGem, \epsVir, \ksiHya, \object{HD220009} and \object{HD107328}. Although the scatter between the nodes is larger than the scatter for dwarfs (see \fig{fig:run01}), it is encouraging to obtain an agreement within 0.08~dex for giants considering the different methods. The mean $1 \sigma$ scatter of all iron abundances for every line is of 0.2~dex, although it is reduced to 0.08~dex when considering only the abundances of the selected lines. FGK giants are challenging objects to model due to their complex atmospheres and large number of lines, in particular lines that form from molecules. In addition, convection in red giants becomes important and 1D models can differ from 3D models,  impacting the final abundances, especially for metal-poor stars \citep{2007A&A...469..687C}. Microturbulence becomes therefore a sensitive parameter, which explains the large error in \vmic\ of \tab{table:BSfeh}.

Typically NLTE departures for this group of stars are negligible when compared with the errors obtained for the abundances, which can be seen in \fig{fig:NTLE-LTE}.  In general, an ionization imbalance of $\sim 0.1$~dex is found for this group of stars, which agrees with the recent conclusion  of \citet{2011ApJ...743..135R}.  The abundances of the selected neutral and ionized iron lines for each giant are shown in \fig{fig:group3}. The dot-dashed lines correspond to the linear regression fits of the \FeI\ abundances as a function of E.P and $\log (\mathrm{EW}/\lambda)$.   While for most of the stars no significant trend of abundances as a function of reduced EW is obtained when considering the error of the fit, a significant positive slope in the regression fit as a function of E.P. is found. The change in abundance over the range in E.P. covered by the lines is however smaller than the final error. 
Thus, we are confident that performing a mean on the abundances  of our selected lines provides robust results for the \feh\ of the benchmark FGK giants.

We obtain typical differences of about $\pm 0.07$~dex or less with the literature values, which is within the uncertainties and scatter found by us and by the literature.  As exceptional  cases, we obtain a slightly lower metallicity of 0.1~dex than the literature value for \ksiHya. The PASTEL catalogue has only two works reporting parameters for this star, where  \citet{1990ApJS...74.1075M} obtained \feh = -0.04 while \citet{Br10} obtained \feh = +0.23. In \tab{table:BMKparams} we present only the latter one due to the restriction on publication year for the extraction from  PASTEL  (see \sect{lit}).  Our value of \feh = 0.12 lies in between those values.  Also,  for \object{HD220009}  we obtain $\sim 0.14$ dex lower than the literature. The only work in  PASTEL after 2000 that reports \feh = -0.67 is that of \citet{2007A&A...468..679S}. The difference can be explained from the different values for the stellar parameters considered by that work, i.e. effective temperature and surface gravity are 100 K and 0.5 dex, respectively, higher than the fundamental values considered by us. 

Finally we comment that during the time when this analysis was carried out by our different groups, we noticed that the effective temperature of \object{HD107328} had been overestimated by 90~K. For that reason, we created a set of line-by-line corrections for \object{HD107328} to account for the lower temperature.  We used the same grid as for the NLTE corrections, but used only LTE curves-of-growth. The uncertainties in the metallicity due to associated errors in the other stellar parameters were then determined using the most recent temperature.

\subsubsection{M giants}\label{supergiants}
The analysis of this group is the most difficult one, where an averaged line-to-line scatter of 0.5~dex is obtained. It includes the stars \psiPhe, \alfCet, \betAra, \gamSge, \alfTau. Note that the spectral class of \alfTau\ is not well established \citep[see ][for a discussion]{lebzelter2012}, being in the limit between late K and early M type. Since our results for \alfTau\  are more comparable to those of the M-type than those of FGK group of giants, for simplicity, we classify \alfTau\ into the M giant group. 

These cool giants have very challenging spectra, mostly because of the presence of molecules. The strength of TiO and CN absorption bands in the coldest stars is particularly high  \citep{1976ApJS...30...61P}, making it extremely difficult to identify the continuum around most of the iron lines. The blends with molecules can become so dominating that an overestimation of metallicity can be obtained when using a given line which has an unidentified molecular blend \citep{1976ApJS...30...61P}.

Additionally, the efficiency of convective energy transport and its effect on line-formation  reaches its maximum at \teff $\sim$ 4000~K \citep{2002A&A...392..619H}. For that reason 3D hydrodynamical models are much more suitable for modeling line-formation in such spectra.  Such models for stars other than the \object{Sun} are not easily available, mainly due to the large computing power needed to model them. In particular, red supergiants  give rise to large granules
that can imprint irregular patterns  \citep{2009A&A...506.1351C, 2010A&A...515A..12C}, but the influence of this effect in spectra of such cool stars has not been investigated so far.  A detailed discussion on spectral modeling for cold giants can be found in \citet{lebzelter2012}. They determined atmospheric parameters of the \bs s \alfCet~and \alfTau~using  11 different methods and made a comparative analysis as for this work. In their analysis (employing also different linelists and atmosphere models between the methods) the unweighted mean values for metallicity were $\mathrm{\feh} = -0.2 \pm 0.2$~dex for both stars. We obtain a value of $-0.45$ for \alfCet\ and $-0.37$ for \alfTau, respectively. Although we obtain values that are more metal-poor, they lie within the errors. 

\begin{figure}
 \resizebox{\hsize}{!}{\includegraphics[angle=90]{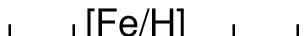}}
  \caption{Trends for group of M giants. }
  \label{fig:group4}
\end{figure}

The abundances of the selected lines can be visualized in \fig{fig:group4}. Because of the reasons explained above, we obtain few un-blended and clean lines that pass our selection criteria. In this work, \alfTau\ and \alfCet\ show a good ionization and excitation balance, although the scatter of the regression fit, as well as the uncertainties of our results are quite high. The other three stars of this group show, on the other hand, a significant slope of the regression fit as a function of E.P.  Note, however, we have no lines at low excitation potentials,  making the regression fit not a good representation of the trend.  We obtain also significant slopes in the regression fit as a function of reduced EW. 


NLTE effects are very small compared with the uncertainties obtained for the abundances. Ionization balance is, on the other hand, unsatisfied for this group except \alfTau\ and \alfCet, when considering the errors. The most extreme cases are \psiPhe\ and \betAra. As discussed in \sect{ionization_balance}, it is impossible to find enough clean and unblended \FeII\ lines in this wavelength domain for such low temperatures, making these \FeII\ determinations thus unrealistic or  not even possible. In the case of \psiPhe, no line passed our selection criteria.  

We recall that we  found only one old reference for metallicity in the PASTEL catalogue for \betAra\ \citep{1979ApJ...232..797L} and no reference for \psiPhe.  Being aware of the difficulties in the analysis of these stars, we expect the \FeI\ abundances obtained by us to be uncertain, but finally  only one of our methods (Porto) could not provide a final value. Given this, we find it encouraging to obtain errors smaller than 0.3 dex and 0.2 dex, for \psiPhe\ and \betAra, respectively. 

\subsubsection{K dwarfs}\label{K}
\begin{figure}
 \resizebox{\hsize}{!}{\includegraphics[angle=90]{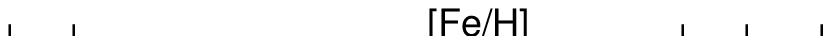}}
  \caption{Trends for group of K dwarfs. }
  \label{fig:group5}
\end{figure}

\Gmb, \cygA, \cygB\ and \epsEri\ are the Benchmark K~dwarfs. As in the previous groups, we plotted in \fig{fig:group5} the abundances of the selected lines for each star in different panels. Even after considering the errors, the members of this group do not present a good excitation balance, since significant trends are obtained for the regression fits for both E.P. and EW. The most extreme case is \cygB, where the \FeI\ abundances increase as a function of E.P. at a rate of $0.107 \pm 0.02$ and decrease as a function of reduced EW at a rate of $0.58 \pm 0.22 $. This star is very cold, and therefore its spectrum is very affected by blends of molecules that are not considered in our line list. A more suitable line list for such cold stars might help in obtaining a better excitation balance.


 61~Cyg~A and 61~Cyg~B belong to a binary system,  therefore the same metallicity for both stars is expected. We obtained a value of -0.33~dex and -0.38~dex for the A and B components, respectively. The difference of 0.05~dex is within the errors. These values are about 0.15~dex lower than the literature values. We attribute this difference to the different temperature adopted  by,  e.g. \citet{2005AJ....129.1063L}, of 4640~K  and 4400~K for the components A and B, respectively. These temperatures are $\sim 300$ K above the values adopted by this work.  Note that \cygB\ does not present a quantification of the ionization balance. Although we could select 2 \FeII\ lines, the mean iron abundance obtained for those lines was of +1.84, which is unphysical. As mentioned above, the reason for such unphysical results comes from the incapacity to detect unblended ionized iron lines for such cool stars.  Thus, we do not list a ionization imbalance or line-by-line standard deviation of \FeII\ lines for \cygB\ in \tab{table:BSfeh_final}.

It is worth mentioning that during one of the first attempts to determine metallicities for this system, the values of fundamental \logg\ considered for the analysis were different (4.49 and 4.61~dex) because they were obtained from evolutionary tracks of \feh\ = -0.10 and \feh = -0.30, for the A and B components of 61~Cyg, respectively.  At that time, we retrieved a new metallicity  of -0.49 and -0.55 dex for 61~Cyg~A and B, respectively, which was translated to a difference in \logg~of -0.06 and -0.08, respectively. A third iteration on \logg\ with the newest metallicity, and a further iteration on \feh\ with the newest surface gravity would be desirable, although we have decided not do to this because of the large errors associated with the mass of this system (see Paper~I) and also the  errors obtained here for the final \feh.

Note that the metallicity obtained for  \Gmb\ differs from the literature by $\sim 0.12$ dex. The 19 works after 2000 in  PASTEL have a mean temperature of  $5090 \pm 89$  K, which is more than 250 K above the fundamental value.   Recently, \citet{2012A&A...545A..17C}, who determined the angular diameter used to obtain the temperature in Paper~I, 
obtained a value that is about 200~K less than the classical spectroscopic values. They suggested a revision of the metallicity based on this fundamental value. We have done this here and we have seen that the consequence is a considerable ionization and excitation imbalance for this new temperature. We have also studied the NTLE effects and concluded that they are not significant in this particular star.  Moreover, \Gmb\ is not so cold as to be strongly affected by molecules, however, its rather low metallicity and mass (0.6~M$_\odot$) suggests 3D or granulation effects caused by convection. The trends found in \fig{fig:group5} and the ionization imbalance could be partly explained by the use of inaccurate 1D LTE models, but we can not exclude the possibility that the fundamental temperature might be too low or perhaps there is another effect that has not been investigated so far, such as magnetic fields or other activity process in the atmosphere. We noted that the knee in the curve of growth of this star is located at a slightly lower reduced equivalent width for than for other stars. This could lead to an inclusion of too strong lines in the EW method , which could bias the \vmic\ measurement and also other slopes in this analysis. The difference in iron abundance due to this effect,  however, should not be as significant as seen here.  If the radius and bolometric flux of this star were accurately determined, then these new stellar parameters would imply that there is a physical process affecting this star in a way that we are not able to quantify. The problem with this hypothesis is that \Gmb\ is a rather ``normal'' star, meaning that it has been commonly studied, and has ``normal'' stellar parameters (not too cool, not too low gravity, not too metal-poor). This makes us believe that  the measurement of its angular diameter might be affected by systematic errors. Until this issue is  resolved, we prefer to point out that this star should be treated with caution as Benchmark Star.

\subsection{Line list: Golden lines}
\label{sect:golden}
In this Section, we give an overview of the \FeI\ and \FeII\ line selection and line data which were used to derive the final metallicity values listed in \tab{table:BSfeh_final}. Only the lines which remained after the selection process described in \sect{errors} were considered.
We determined which lines were used in common for each of the groups defined in Section~\ref{discussion} and refer to these as the ``golden lines''. We found that there were significant differences in line selection between individual methods within several star groups, and thus the group definitions were somewhat expanded as explained below.
The unique lists of 171 \FeI\ and 13 \FeII\ lines occurring in any of the groups can be found in Tables~\ref{tab:golden_Fe1} and ~\ref{tab:golden_Fe2}, respectively. The tables give the most relevant atomic data. For the lines identified for each individual group, we give the minimum and maximum standard deviations of the average line abundances, and the minimum and maximum number of abundances averaged for each line in the respective column.

The \emph{metal-poor stars} (\sect{metal-poor}) were divided into dwarfs (\object{HD84937}, \object{HD140283}) and giants (\object{\object{HD122563}}) and designated ``MPD'' and ``MPG'', respectively, in Tables~\ref{tab:golden_Fe1} and ~\ref{tab:golden_Fe2}. As can be seen in the tables, the number of golden lines is considerably larger for the metal-poor giant (56 \FeI\ and 4 \FeII\ lines) than for metal-poor dwarfs (17 \FeI\ and 1 \FeII\ lines). 15 of the MPD \FeI\ lines are contained in the MPG list, while the single \FeII\ line common to the MPDs is different from the MPG \FeII\ lines.
For the two MPDs, the standard deviations of the abundances are rather similar for all lines.

The group of \emph{FG~dwarfs} (\sect{FG}) contains four stars for which the \FeI\ line selection differs from the others. This sub-group is designated ``FGDb'' in \tab{tab:golden_Fe1} and comprises  \etaBoo, \object{HD22879}, \object{HD49933}, and \Procyon. The remaining stars listed in \sect{FG} are designated ``FGDa''.
In general, the metallicity of the stars in the FGDb group is based on fewer \FeI\ lines than those in FGDa (see \tab{table:BSfeh_final}). However, the number of golden \FeI\ lines is similar for FGDa and FGDb (79 and 74, respectively), with 51 lines in common between the two sub-groups. The four stars in FGDb differ from those in FGDa in various respects, which reduce the number of useful lines: \object{HD49933} and \Procyon\ have the highest effective temperatures, \object{HD49933} and \etaBoo\ have the largest \vsini, and \object{HD22879} is a moderately metal-poor star.
The \FeII\ line lists are more homogeneous, resulting in six golden lines for all stars, with two exceptions as noted in Tab.~\ref{tab:golden_Fe2} (column ``FGD'').
For the FG~dwarfs, the abundance dispersions show a large variation from star to star.
A more detailed investigation for the FGDa group shows that for \FeI\ lines the lowest minimum values are mostly due to the star \betHyi, and the highest maximum values to \betVir.
For most other stars and most \FeI\ lines the dispersion is around 0.06~dex.
In the FGDb group, \Procyon\ and \object{HD22879} have the minimum dispersion for half of the \FeI\ lines each (and \object{HD49933} for seven lines). 
The maximum dispersion is mostly due to \etaBoo\ (61 lines), and sometimes to \object{HD49933} or \Procyon\ (12 and 1 lines, respectively).
The mode of the dispersion for FGDs is about 0.04~dex for all \FeII\ lines.

The group of \emph{FGK~giants} consists of the stars listed in \sect{RG}. The 101 golden \FeI\ and 6 \FeII\ lines identified for this group are marked in column ``FGKG'' in Tables~\ref{tab:golden_Fe1} and ~\ref{tab:golden_Fe2}.
For this group, the variation of dispersions is even larger than for FG~dwarfs.
The maximum dispersion for \FeI\ lines is mainly seen for \muLeo\ (82 lines), while the minimum dispersion occurs mainly for \object{HD220009}, \object{HD107328}, and \Arcturus\ (for 54, 16, and 12 lines respectively).
For most other stars, the dispersion scatters around 0.08~dex.
Also for the \FeII\ lines, the largest dispersion is found for \muLeo.
The dispersion is in general higher than for FG~dwarfs (around 0.12~dex).

The group of \emph{M~giants} consists of the stars listed in \sect{supergiants}, with one exception. The line list for \psiPhe\ differs significantly from the other stars (23 \FeI\ lines, of which only 6 are in common with the others). The 21 golden \FeI\ and 3 \FeII\ lines identified for this group are marked in column ``MG'' in Tables~\ref{tab:golden_Fe1} and ~\ref{tab:golden_Fe2}, while \psiPhe\ is listed in a separate column in Table~\ref{tab:golden_Fe1} (no \FeII\ lines were selected for this star). 
The minimum abundance dispersion for \FeI\ lines in M~giants is mostly found for \gamSge\ (13 lines), and the maximum dispersion equally often in \alfCet\ and \betAra\ (7 and 8 lines, respectively).
\psiPhe\ shows in general high dispersions, with the notable exceptions of the \FeI\ lines at 6219.28 and 6336.82~\AA, with dispersions of about 0.1~dex.

Finally, the group of \emph{K~dwarfs} described in \sect{K} was divided into two sub-groups with two different lists of golden \FeI\ lines. These are designated ``KDa'' (\cygA, \epsEri) and ``KDb'' (\cygB, \Gmb) in Table~\ref{tab:golden_Fe1}, with 127 and 85 \FeI\ lines, respectively, and 72 lines in common between the two sub-groups.
The differences in line selection between the two sub-groups may be related to the specific parameter combinations (\teff,\feh) of the stars.
In the KDa group, the maximum dispersion occurs for \cygA\ for 2/3 of the lines.
In the KDb group, \cygB\ accounts for the maximum dispersion for most of the lines (77).
Regarding the \FeII\ lines, the star \epsEri\ stands out among the group members, with the largest number of lines selected (11 compared to 1--4). These are marked in Table~\ref{tab:golden_Fe2} in column ``KD'', which includes a note identifying the lines in common with the other three stars.

\subsubsection{Discrepant lines}

It is important to discuss here that while selecting the golden lines, we found that  in some cases the derived abundances by our methods differed significantly, i.e. up to 0.4~dex even for FG dwarfs for which we obtain the lowest line-to-line scatter in the final abundance determination (see above). This was surprising, since our golden lines were chosen to be unblended and are located in spectral regions with easy continuum placement. Moreover, our analysis is based on a great effort to have  homogeneous atomic data and  atmospheric models, making such differences difficult to explain. 

Thus, we made a deep investigation of this issue and considered 4 examples of discrepant lines. This analysis was carried out mainly by M. Bergemann, U. Heiter, P. Jofr\'e, K. Lind, T. Masseron, J. Sobeck and  H. Tabernero. We compared three of  the radiative transfer codes (SME, MOOG, Turbospectrum) and found that their profiles were consistent when considering the same stellar parameters (they were set to those of the \object{Sun}, \Arcturus, \object{HD84937}, and \object{HD140283}). Naturally, a difference could still be seen due to different prescriptions and treatment of lines and spectrum formation (collisional broadening, radiative broadening,
scattering, limb darkening, spherical geometry, to name a few). But all together, this did not explain the 0.4~dex of the discrepant line examples. 

We concluded that these discrepancies come apparently from a combination of different measured equivalent widths (differing up to 60\%), the details of the fitting procedures, the choice of microturbulence parameter (see \fig{fig:run01_vmic} for the different values) and the continuum placement. Understanding the contribution in the final discrepancy of each individual line from each of the aforementioned sources goes beyond the purpose of this paper. Here, we aim to combine abundances of numerous lines and methods homogeneously and provide a reference value for the metallicity of Benchmark Stars.  In general, our results agree very well on a line-by-line basis, and cases such as those discussed here are rare. However, we point out that this problem can arise even after performing analyses focused on homogeneity. Therefore, it is worthwhile to investigate further the sources of these discrepancies.

\subsection{The final metallicity and its uncertainties}

We have extensively mentioned in this manuscript that our revision of metallicity, using fixed \teff\ and \logg\ with values that are independent from spectroscopic analysis, does not necessarily give a  \feh\ value that satisfies ionization and excitation balance. The differences can be used as a way to quantify the uncertainties in the theoretical assumptions used to compute synthetic line profiles and atmospheric models. This reminds us  the motivation of defining a set of Benchmark Stars: the importance of having a standard set of stars with stellar parameters that are independent from spectroscopy, as this helps to make improvements to  spectroscopic methods and models for stars of different spectral types. 

It is worth discussing that in this work we obtained three different values for metallicity: a value from \FeI\ lines, another one from \FeII\  lines and a final one from \FeI\ lines after corrections from NLTE effects. Since we aimed to provide a reference metallicity, we chose the metallicity from \FeI\ lines after NLTE corrections to be our final value. Although we know that neutral iron lines are more sensitive to NLTE effects, we do not obtain a metallicity from ionized iron lines for all stars. Since we aim for  homogeneity in this work, we prefer to consider the results from neutral lines for the final value. Finally, we have two main reasons in favour of choosing  the NLTE values:  (i) that the ionisation balance is slightly restored after NLTE corrections and (ii) it is more accurate. 

The natural question that arises from our choice is of how to use the \feh\ of this work when someone has a parametrisation method that employes LTE, or that obtains \FeII\ abundance. The answer to this question is that we consider these values as part of the uncertainties of the final value. In other words, we quantify here the error associated with the ionization imbalance and with NLTE effects by providing the difference between the final value and  [\FeII/H]  or [\FeI/H] with LTE approximation, respectively.  These values are found in \tab{table:BSfeh_final} labeled as $\Delta$(ion) and $\Delta$ (LTE). To retrieve the metallicity value that one would obtain using \FeII\ lines, one needs to determine $\mathrm{\feh} - \Delta \mathrm{(ion)}$. Similarly, to retrieve the value obtained under LTE, one needs to determine $\mathrm{\feh} - \Delta \mathrm{(LTE)}$.  Moreover, the full information of  iron abundances and EWs for each selected line and individual method can be retrieved from the online tables. 

To finish this section, we summarise that our reference  metallicity is the one obtained by averaging the NLTE abundances of the selected \FeI\ lines. This value is associated with a series of sources of uncertainties, which are (i) the scatter in the line-by-line analysis of the selected lines; (ii) the difference in \feh when considering the uncertainty in the fundamental \teff; (iii) the difference in \feh when considering the uncertainty in the fundamental \logg; (iv) the difference in \feh when considering the uncertainty in \vmic, (v) the difference in \feh obtained from neutral and ionised iron lines; and (vi) the difference in \feh obtained from LTE and NLTE analyses. The final value and its 6 sources of errors are listed in \tab{table:BSfeh_final}. In addition, the line-by-line standard deviation from \FeII\ abundances and the number of lines employed for the determination of metallicity from neutral and ionized iron lines are also indicated in the Table.

\section{Summary and Conclusions}\label{conclusions}

We have made an extensive study on the determination of metallicity for the sample of 34 FGK \bs s introduced in Paper~I. In this study we performed a spectral analysis of high SNR and high resolution ($R \geq 70000$) spectra taken from the library of \bs s described in Paper~II. Two different libraries were analyzed, one with the spectra at their original resolution and the other one convolved to $R = 70000$. In addition, the  analysis was done for the same star observed with different instruments. 

The analysis consisted of fixing effective temperature and surface gravity to the fundamental values presented in Paper~I, and determining metallicity and microturbulence velocity simultaneously. Up to seven different methods were used for this analysis, all of them considering the same input material, such as spectra, line list and atmosphere models. 

Three different runs were performed: {\it run-nodes}, consisting in the analysis of one spectrum per \bs,  that allows a one-to-one comparison between different methods; {\it run-resolutions}, consisting in the analysis as in the same spectrum of the previous run, but using this time its version in original resolution. This run allowed the study of the impact of the varied resolution. The third run, {\it run-instruments} consisted in the analysis of the whole library convolved to R = 70000, and allowed us to study instrumental effects. We obtained consistent and robust results, where the final metallicity was not biased either by method, resolution nor instrument. 

Since we fixed \teff\ and \logg\ by values that are independent of spectroscopy, the metallicity analysis resulted in \FeI\ and \FeII\ abundances that did not necessarily agree. The comparison between neutral and ionized iron abundances was discussed, together with a quantification of how much  \teff\ and \logg\  would need to deviate from the fundamental value in order to comply  with ionization balance, excitation balance and line strength balance. This was done by a test of determining \teff, \logg\ and \vmic\ together with \feh. 

To provide a final value of metallicity, we combined our results using a line-by-line approach. Starting from all individual abundances of every method, we selected only those lines which were analyzed by at least three methods and agreed within $2 \sigma$ with the average abundance calculated from all lines. The selected lines were then averaged to have only one abundance per line, which was then used to perform NLTE corrections and quality checks such as ionization and excitation balance.  Our final value consists of the iron abundance obtained from \FeI\ lines after NLTE corrections.

 We studied many  different sources of errors, which are all reported separately. The first one comes from the consideration of the $1 \sigma$ scatter of the line-by-line analysis. Then, we determined the uncertainty of the metallicity due to the errors associated with the effective temperature, surface gravity and microturbulent velocity. To do so, iron abundances were calculated by performing 6 additional runs only on the selected lines, each run fixing \teff, \logg\ and \vmic\ to the values considering their associated errors. Finally,  errors due to ionization imbalance and deviations from NLTE were quantified. 

Generally, we were able to obtain robust values for \feh\ for the stars of our sample, making this work the first one to determine metallicity homogeneously for the complete set of Gaia FGK \bs s. Our final \feh\ values are thus appropriate for use as reference values.  When comparing our results with previous studies in the literature, we obtain a good agreement for 28 stars and different values for 4 stars (\object{HD220009}, \cygA, \cygB, \betAra), which we adopt as a new reference \feh. In addition, we provide for the first time a value for the metallicity of \psiPhe.   Although we obtain very different metallicities for \Gmb\ compared to the literature, we prefer to  caution against defining a new set of reference parameters for this star, as we are unable to understand the reason for this discrepancy and further investigations on its fundamental parameters are needed.  The final reference values and their uncertainties are indicated in \tab{table:BSfeh_final}. Having well determined stellar parameters for the \bs s will improve the homogeneous analyses of current stellar surveys, which have become a key piece in Galactic studies. 

We made a careful study in the selection of  candidates to serve as benchmarks for stellar spectra analyses. The accurate distance and angular diameter of these stars provide us with fundamental determinations of effective temperature and surface gravity. Their proximity and brightness provide us with the possibility of having high quality spectra that are suitable for a more precise determination of metallicity. This paper on the series of Gaia FGK Benchmark Stars, together with Paper~I and Paper~II, describe and discuss extensively our choice for the reference values of the three main stellar parameters \teff, \logg\ and \feh.  We encourage our colleagues to use the spectra of the Gaia \bs\ and their parameters to evaluate the performance of parametrization methods as a way to relate the data to the Gaia-ESO Survey.  We can transform our spectra such that they look like the data taken from other spectrographs, and our metallicities can be reproduced as we document in the online tables each individual value used for its final determination. Using this material will allow for the connection of different methods and cross-calibration of surveys, leading to a more consistent understanding of the structure and evolution of our Galaxy.

\begin{acknowledgements} 
We thank all LUMBA members for the rich discussions on the development of SME for automatic analyses of spectra, which were crucial to the development of the setups used for the analysis of this project.  P.J. acknowledges the useful comments and proof reading done by T. M\"adler. U.H. acknowledges support from the Swedish National Space Board (Rymdstyrelsen). S.G.S acknowledges the support from the Funda\c{c}\~ao para a Ci\^encia e Tecnologia (Portugal) in the form of the grants SFRH/BPD/47611/2008. The computations for the AMBRE project have been performed with the high-performance computing facility SIGAMM, hosted by OCA.  R.S. acknowledges the support of the ASI (Agenzia Spaziale Italiana) under contracts to
INAF I/037/08/0 and I/058/10/0. Finally, P.J. and U.H. acknowledge the contribution from the suggestions of the anonymous referee. 
\end{acknowledgements}

\small

\bibliographystyle{aa} 
\bibliography{references_tot} 
\normalsize

\onecolumn
\longtab{4}{
\small
\begin{landscape}
\centering
\begin{longtable}{rrrrrccccccccc}
\caption{ List of ``golden'' Fe~I lines for various groups of stars (see text for definition of groups). \label{tab:golden_Fe1}} \\
\hline\hline
$\lambda$ [\AA] & E [eV] & $\log gf$ &    Waals & Ref & MPD  (2) & MPG  (1) & FGDa (11) & FGDb (4) & FGKG (7) & MG   (4) & $\psi$~Phe & KDa  (2) & KDb  (2)\\
\hline
\endfirsthead
\caption{ continued.} \\
\hline\hline
$\lambda$ [\AA] & E [eV] & $\log gf$ &    Waals & Ref & MPD  (2) & MPG  (1) & FGDa (11) & FGDb (4) & FGKG (7) & MG   (4) & $\psi$~Phe & KDa  (2) & KDb  (2)\\
\hline
\endhead
\hline
\endfoot
\hline\hline
\multicolumn{14}{l}{\parbox{1.00\textheight}{Column descriptions: $\lambda$: wavelength. $E_{\rm low}$: lower level energy. ``Waals'': parameters used to calculate line broadening due to collisions with neutral hydrogen; integer part: broadening cross-section at a velocity of 10$^4$~m s$^{-1}$ in atomic units, fractional part: velocity parameter \citep[see][]{BPM}; if zero, the Uns\"old approximation was used. ``Ref'': reference code for the $gf$-values (see below). The remaining columns are headed by a label for each group defined in the text, and the number of stars in parantheses. The columns give, for each group, the minimum and maximum standard deviations of the average line abundances, and the minimum and maximum number of abundances averaged for each line.}}\\
\multicolumn{14}{l}{\parbox{1.00\textheight}{References: 102: \citet{1991AaA...248..315B, 1994AaA...282.1014B, 1979MNRAS.186..633B, 1979MNRAS.186..657B, 1982MNRAS.199...43B, 1982MNRAS.201..595B, 1995AaA...296..217B, 1991JOSAB...8.1185O}. 114: \citet{1974ApJ...192..793B, 2006JPCRD..35.1669F}.  129: \citet{1969AaA.....2..274G, 1988JPCRD..17S....F}.  156: \citet{1974AaAS...18..405M}.  167: \citet{1970AaA.....9...37R, 1988JPCRD..17S....F}.  186: \citet{1970ApJ...162.1037W, 1988JPCRD..17S....F}.  187: \citet{1971ApJ...166L..31W, 1988JPCRD..17S....F}. }}\\
\endlastfoot
         4787.83 &             2.9980 &    -2.563 &  818.227 & 102 &      &      &      &      & 0.04/0.41 3/3 &      &      & 0.06/0.07 3/3 &      \\
         4788.76 &             3.2370 &    -1.763 &  238.249 & 102 &      &      & 0.02/0.16 3/3 & 0.03/0.09 3/3 & 0.06/0.52 3/3 &      &      & 0.05/0.11 3/3 &      \\
         4802.88 &             3.6420 &    -1.514 &  356.244 & 102 &      &      &      &      &      &      &      & 0.11/0.30 3/4 &      \\
         4808.15 &             3.2510 &    -2.690 &  297.274 & 156 &      &      & 0.04/0.23 5/6 & 0.04/0.16 3/5 & 0.06/0.31 5/6 & 0.14/0.25 3/4 &      & 0.03/0.04 6/6 & 0.05/0.06 4/5 \\
         4869.46 &             3.5460 &    -2.420 &  246.248 & 156 &      &      &      &      & 0.06/0.65 3/3 &      &      &      &      \\
         4875.88 &             3.3320 &    -1.920 &  848.231 & 156 &      &      & 0.07/0.39 4/4 & 0.08/0.21 3/4 & 0.04/0.51 4/4 &      &      & 0.13/0.15 3/4 & 0.08/0.21 3/3 \\
         4877.60 &             2.9980 &    -3.050 &  795.230 & 156 &      &      &      &      & 0.04/0.30 3/3 &      &      & 0.05/0.07 3/3 &      \\
         4907.73 &             3.4300 &    -1.840 &  909.227 & 129 &      &      & 0.04/0.14 3/3 &      & 0.02/0.40 3/3 &      &      & 0.07/0.09 3/3 &      \\
         4924.77 &             2.2790 &    -2.178 &  360.244 & 102 &      & 0.04 3 &      & 0.03/0.10 3/3 &      &      &      &      & 0.11/0.11 3/3 \\
         4946.39 &             3.3680 &    -1.170 &  848.232 & 187 &      &      &      & 0.04/0.13 3/3 &      &      &      &      &      \\
         4950.11 &             3.4170 &    -1.670 &  880.228 & 129 &      & 0.02 4 & 0.07/0.22 3/5 & 0.03/0.14 4/5 & 0.06/0.16 3/4 &      &      & 0.14/0.16 4/4 & 0.05/0.16 3/5 \\
         4962.57 &             4.1780 &    -1.182 &    0.000 & 102 &      &      & 0.02/0.19 3/4 & 0.01/0.15 3/4 &      &      &      & 0.01/0.12 4/4 & 0.01/0.16 3/4 \\
         4969.92 &             4.2170 &    -0.710 &  962.279 & 129 &      &      &      & 0.01/0.16 3/3 &      &      &      &      & 0.03/0.10 3/3 \\
         4985.55 &             2.8650 &    -1.340 &  727.238 & 190 &      & 0.03 3 &      &      &      &      &      &      &      \\
         4994.13 &             0.9150 &    -3.002 &  246.245 & 102 & 0.03/0.05 3/3 & 0.04 3 &      & 0.08/0.18 3/4 &      &      &      &      &      \\
         5001.86 &             3.8810 &    -0.010 &  725.240 & 114 &      & 0.02 3 &      &      &      &      &      &      &      \\
         5012.69 &             4.2830 &    -1.690 & 1020.279 & 156 &      &      & 0.05/0.20 3/4 &      & 0.07/0.30 3/4 &      &      & 0.06/0.11 4/4 &      \\
         5044.21 &             2.8510 &    -2.038 &  713.238 & 102 &      &      &      & 0.04/0.11 3/3 &      &      &      &      &      \\
         5049.82 &             2.2790 &    -1.349 &  353.239 & 102 &      & 0.01 3 &      &      &      &      &      &      &      \\
         5058.50 &             3.6420 &    -2.830 &  353.313 & 167 &      &      & 0.09/0.20 3/5 &      & 0.09/0.22 4/5 & 0.15/0.40 3/4 &      & 0.12/0.28 4/4 &      \\
         5060.08 &             0.0000 &    -5.431 &    0.000 & 102 &      &      &      &      &      &      &      & 0.17/0.29 3/3 & 0.08/0.08 3/3 \\
         5088.15 &             4.1540 &    -1.680 &  810.278 & 156 &      &      &      &      & 0.09/0.52 3/3 &      &      &      &      \\
         5107.45 &             0.9900 &    -3.091 &  248.245 & 102 &      & 0.04 3 &      &      &      &      &      &      &      \\
         5107.64 &             1.5570 &    -2.358 &  289.258 & 102 &      & 0.02 3 &      &      &      &      &      &      &      \\
         5109.65 &             4.3010 &    -0.980 &  980.280 & 167 &      &      &      &      &      &      &      & 0.07/0.15 3/3 & 0.04/0.40 3/3 \\
         5127.36 &             0.9150 &    -3.278 &  243.246 & 102 &      & 0.03 3 &      & 0.05/0.16 3/3 &      &      &      &      &      \\
         5131.47 &             2.2230 &    -2.515 &  356.274 & 102 &      &      &      & 0.03/0.17 3/3 &      &      &      & 0.11/0.34 3/3 & 0.09/0.09 3/3 \\
         5141.74 &             2.4240 &    -2.101 &  367.251 & 102 &      & 0.09 4 & 0.03/0.37 3/5 & 0.04/0.24 4/5 &      &      &      & 0.08/0.20 3/4 &      \\
         5194.94 &             1.5570 &    -2.021 &  286.255 & 102 &      & 0.07 3 &      &      &      &      &      &      &      \\
         5197.94 &             4.3010 &    -1.540 &  925.279 & 156 &      &      & 0.06/0.20 4/5 & 0.04/0.14 4/4 & 0.17/0.57 4/5 &      &      & 0.03/0.19 3/5 &      \\
         5198.71 &             2.2230 &    -2.113 &  351.271 & 102 &      & 0.03 3 &      & 0.03/0.21 3/3 &      &      &      &      &      \\
         5215.18 &             3.2660 &    -0.871 &  849.229 & 102 &      & 0.03 3 &      &      &      &      &      &      & 0.03/0.05 3/3 \\
         5217.39 &             3.2110 &    -1.116 &  815.232 & 102 & 0.03/0.07 3/3 & 0.05 4 &      & 0.07/0.18 3/4 &      &      &      &      & 0.09/0.23 3/4 \\
         5223.18 &             3.6350 &    -1.783 &  390.253 & 102 &      &      &      &      & 0.05/0.34 3/3 &      &      &      &      \\
         5225.53 &             0.1100 &    -4.755 &  207.253 & 102 &      & 0.04 4 &      & 0.04/0.27 3/4 &      &      &      &      & 0.11/0.21 4/4 \\
         5228.38 &             4.2200 &    -1.190 &  809.278 & 156 &      &      &      &      &      &      &      & 0.07/0.22 3/3 &      \\
         5232.94 &             2.9400 &    -0.076 &  713.238 & 102 & 0.06/0.07 5/5 & 0.08 4 &      &      &      &      &      &      &      \\
         5242.49 &             3.6340 &    -0.967 &  361.248 & 102 &      & 0.07 3 &      & 0.05/0.10 3/4 &      &      &      & 0.12/0.23 3/4 & 0.06/0.06 3/4 \\
         5243.78 &             4.2560 &    -1.050 &  842.278 & 156 &      &      &      & 0.00/0.25 4/4 & 0.13/0.46 3/4 &      &      & 0.07/0.09 4/4 & 0.04/0.05 3/4 \\
         5247.05 &             0.0870 &    -4.975 &  206.253 & 102 &      & 0.04 3 &      &      &      &      &      & 0.02/0.13 3/3 & 0.15/0.33 3/3 \\
         5250.21 &             0.1210 &    -4.918 &  207.253 & 102 &      & 0.03 3 &      &      &      &      &      & 0.13/0.20 3/3 & 0.10/0.10 3/3 \\
         5250.65 &             2.1980 &    -2.180 &  344.268 & 102 & 0.06/0.07 3/3 & 0.01 4 &      & 0.04/0.13 3/4 & 0.09/0.26 3/3 &      &      & 0.24/0.27 3/3 & 0.11/0.15 3/4 \\
         5253.02 &             2.2790 &    -3.840 &  368.253 & 156 &      &      & 0.03/0.13 4/5 &      & 0.06/0.26 4/5 &      & 0.51 3 & 0.06/0.09 5/5 &      \\
         5253.46 &             3.2830 &    -1.573 &  849.229 & 102 &      &      &      & 0.04/0.20 3/3 &      &      &      & 0.17/0.19 3/3 &      \\
         5285.13 &             4.4340 &    -1.540 & 1046.282 & 156 &      &      &      &      & 0.05/0.25 3/3 &      &      & 0.02/0.09 3/3 &      \\
         5288.52 &             3.6940 &    -1.508 &  353.297 & 102 &      &      &      &      &      &      &      & 0.06/0.08 3/3 &      \\
         5293.96 &             4.1430 &    -1.770 &    0.000 & 156 &      &      & 0.05/0.15 3/4 & 0.05/0.12 3/4 & 0.04/0.23 3/4 &      &      & 0.03/0.04 4/4 &      \\
         5294.55 &             3.6400 &    -2.760 &  394.237 & 156 &      &      & 0.04/0.41 4/5 &      & 0.05/0.17 5/5 &      &      & 0.04/0.22 3/5 &      \\
         5295.31 &             4.4150 &    -1.590 & 1014.281 & 156 &      &      & 0.03/0.19 3/5 & 0.02/0.11 4/5 & 0.04/0.21 5/5 &      &      & 0.05/0.09 5/5 & 0.06/0.18 3/3 \\
         5302.30 &             3.2830 &    -0.720 &  835.231 & 102 &      & 0.07 4 &      &      &      &      &      &      &      \\
         5321.11 &             4.4340 &    -1.089 & 1024.281 & 102 &      &      &      &      &      &      &      & 0.03/0.10 3/3 &      \\
         5322.04 &             2.2790 &    -2.802 &  341.236 & 102 &      &      &      &      &      &      &      & 0.07/0.10 3/3 & 0.02/0.04 3/3 \\
         5339.93 &             3.2660 &    -0.684 &  815.234 & 102 &      & 0.06 4 &      &      &      &      &      &      &      \\
         5365.40 &             3.5730 &    -1.020 &  283.261 & 102 &      & 0.15 3 &      & 0.05/0.19 3/4 &      &      & 0.39 3 & 0.08/0.09 3/4 & 0.06/0.08 3/4 \\
         5367.47 &             4.4150 &     0.444 &  972.280 & 102 &      & 0.08 3 &      &      &      &      &      &      &      \\
         5373.71 &             4.4730 &    -0.760 & 1044.282 & 156 &      &      & 0.03/0.12 3/4 & 0.00/0.08 3/4 & 0.06/0.27 3/4 &      &      & 0.09/0.18 4/4 &      \\
         5379.57 &             3.6940 &    -1.514 &  363.249 & 102 &      &      & 0.01/0.13 3/4 & 0.02/0.25 3/4 &      &      &      & 0.06/0.08 4/4 & 0.01/0.11 3/4 \\
         5386.33 &             4.1540 &    -1.670 &  930.278 & 156 &      &      & 0.06/0.21 3/5 & 0.04/0.16 4/5 & 0.06/0.30 5/5 &      & 0.66 3 & 0.08/0.13 5/5 & 0.07/0.22 3/4 \\
         5389.48 &             4.4150 &    -0.410 &  959.280 & 187 &      & 0.06 3 &      & 0.02/0.07 3/4 &      &      &      & 0.08/0.09 3/4 & 0.02/0.04 3/4 \\
         5395.22 &             4.4450 &    -2.070 &  995.281 & 156 &      &      &      &      & 0.03/0.19 3/4 &      & 0.25 3 & 0.09/0.09 3/3 &      \\
         5397.13 &             0.9150 &    -1.988 &  238.249 & 102 & 0.10/0.13 3/3 &      &      &      &      &      &      &      &      \\
         5398.28 &             4.4450 &    -0.630 &  993.280 & 156 &      &      & 0.04/0.10 3/5 & 0.01/0.29 4/5 & 0.07/0.31 3/5 &      & 0.49 3 & 0.10/0.16 4/4 & 0.02/0.04 3/5 \\
         5412.78 &             4.4340 &    -1.716 &  971.280 & 102 &      &      & 0.01/0.12 3/4 &      & 0.03/0.18 4/4 &      & 0.17 3 & 0.03/0.05 3/4 &      \\
         5415.20 &             4.3860 &     0.643 &  910.279 & 102 &      & 0.09 3 &      &      &      &      &      &      &      \\
         5417.03 &             4.4150 &    -1.580 &  944.280 & 156 &      &      & 0.04/0.30 4/5 & 0.01/0.11 3/5 & 0.05/0.30 4/5 &      &      & 0.06/0.14 5/5 & 0.13/0.59 3/4 \\
         5424.07 &             4.3200 &     0.520 &  825.278 & 186 & 0.03/0.06 3/3 & 0.02 3 &      &      &      &      &      &      &      \\
         5434.52 &             1.0110 &    -2.119 &  243.247 & 102 & 0.09/0.22 3/3 &      &      &      &      &      &      &      &      \\
         5441.34 &             4.3120 &    -1.630 &  807.278 & 156 &      &      & 0.02/0.25 4/5 &      & 0.06/0.32 4/5 & 0.17/0.39 3/4 &      & 0.05/0.18 5/5 & 0.10/0.11 3/3 \\
         5445.04 &             4.3860 &    -0.020 &  895.279 & 186 &      &      &      &      &      &      &      &      & 0.03/0.04 3/3 \\
         5464.28 &             4.1430 &    -1.402 &  380.250 & 102 &      &      &      &      & 0.06/0.33 3/3 &      &      & 0.07/0.08 3/3 &      \\
         5466.40 &             4.3710 &    -0.630 &  865.278 & 187 &      &      &      & 0.02/0.13 3/4 & 0.14/0.45 4/4 &      &      & 0.13/0.29 4/4 & 0.03/0.09 4/4 \\
         5470.09 &             4.4460 &    -1.710 &  953.280 & 156 &      &      & 0.05/0.13 3/4 &      & 0.05/0.25 3/4 &      &      & 0.07/0.09 3/4 & 0.08/0.12 3/3 \\
         5473.90 &             4.1540 &    -0.790 &  738.241 & 114 &      &      &      &      &      &      &      & 0.12/0.16 3/3 & 0.01/0.03 3/3 \\
         5483.10 &             4.1540 &    -1.406 &  737.241 & 102 &      &      &      &      &      &      &      & 0.06/0.08 3/3 &      \\
         5487.15 &             4.4150 &    -1.430 &  908.279 & 156 &      &      &      &      & 0.13/0.37 3/3 &      &      &      &      \\
         5494.46 &             4.0760 &    -1.990 &    0.000 & 156 &      &      & 0.01/0.24 3/4 &      & 0.08/0.33 4/4 &      &      &      &      \\
         5522.45 &             4.2090 &    -1.450 &  744.215 & 156 &      &      & 0.02/0.09 3/4 & 0.02/0.09 3/4 & 0.06/0.31 4/4 & 0.06/0.23 3/4 & 0.27 3 & 0.05/0.05 4/4 & 0.01/0.13 3/4 \\
         5539.28 &             3.6420 &    -2.560 &  383.260 & 156 &      &      & 0.03/0.35 3/4 &      & 0.08/0.29 3/4 &      &      & 0.02/0.16 3/4 &      \\
         5543.94 &             4.2170 &    -1.040 &  742.238 & 156 &      &      & 0.01/0.10 3/4 & 0.02/0.14 3/4 & 0.04/0.33 3/4 &      & 0.22 3 & 0.06/0.06 4/4 & 0.01/0.21 4/4 \\
         5546.51 &             4.3710 &    -1.210 &  825.278 & 156 &      &      & 0.05/0.10 3/4 & 0.04/0.17 3/4 & 0.06/0.33 3/4 &      &      & 0.05/0.06 4/4 &      \\
         5560.21 &             4.4340 &    -1.090 &  895.278 & 156 &      &      & 0.03/0.10 3/4 & 0.03/0.20 3/4 & 0.06/0.28 3/4 & 0.02/0.20 3/4 &      & 0.05/0.07 4/4 & 0.01/0.01 3/3 \\
         5569.62 &             3.4170 &    -0.486 &  848.233 & 102 & 0.03/0.03 3/3 & 0.08 4 &      &      &      &      &      &      &      \\
         5576.09 &             3.4300 &    -0.900 &  854.232 & 156 & 0.03/0.06 3/3 & 0.06 4 &      & 0.06/0.15 3/4 &      &      &      &      &      \\
         5586.76 &             3.3680 &    -0.120 &  817.238 & 102 &      & 0.02 3 &      &      &      &      &      &      &      \\
         5618.63 &             4.2090 &    -1.275 &  732.214 & 102 &      &      & 0.02/0.09 5/5 & 0.02/0.21 4/5 & 0.04/0.24 4/5 &      &      & 0.05/0.08 5/5 &      \\
         5619.60 &             4.3860 &    -1.600 &  808.277 & 156 &      &      & 0.02/0.26 3/4 & 0.01/0.05 3/3 & 0.04/0.41 4/4 &      &      & 0.03/0.04 4/4 &      \\
         5633.95 &             4.9910 &    -0.230 &  635.270 & 156 &      &      & 0.02/0.18 3/4 & 0.01/0.18 3/4 & 0.08/0.42 3/4 &      &      & 0.04/0.07 4/4 & 0.03/0.13 3/4 \\
         5636.70 &             3.6400 &    -2.510 &  368.310 & 156 &      &      & 0.02/0.33 3/4 &      & 0.02/0.27 4/4 &      &      & 0.03/0.04 4/4 &      \\
         5638.26 &             4.2200 &    -0.770 &  730.235 & 156 &      &      & 0.03/0.17 4/5 & 0.03/0.17 4/5 & 0.09/0.41 4/5 &      &      & 0.05/0.09 4/5 & 0.02/0.13 5/5 \\
         5641.43 &             4.2560 &    -1.080 &  739.234 & 156 &      &      & 0.01/0.28 3/3 &      &      &      &      & 0.03/0.05 3/3 &      \\
         5649.99 &             5.0990 &    -0.820 &  719.265 & 156 &      &      &      &      & 0.04/0.24 3/3 &      &      & 0.02/0.04 3/3 &      \\
         5651.47 &             4.4730 &    -1.900 &  898.278 & 156 &      &      & 0.04/0.17 5/6 &      & 0.04/0.18 6/6 & 0.12/0.20 3/4 &      & 0.04/0.05 5/6 &      \\
         5652.32 &             4.2600 &    -1.850 &  754.210 & 156 &      &      & 0.05/0.22 4/5 &      & 0.02/0.27 4/5 &      &      & 0.04/0.04 5/5 & 0.10/0.28 3/3 \\
         5653.87 &             4.3860 &    -1.540 &  792.277 & 156 &      &      & 0.01/0.13 3/3 &      & 0.01/0.14 3/3 &      &      & 0.02/0.10 3/3 & 0.04/0.04 3/3 \\
         5655.18 &             5.0640 &    -0.600 &    0.000 & 156 &      &      &      &      &      &      &      & 0.08/0.27 3/3 &      \\
         5661.35 &             4.2840 &    -1.756 &  765.209 & 102 &      &      & 0.03/0.10 3/4 &      & 0.02/0.20 3/4 &      &      & 0.04/0.08 4/4 &      \\
         5662.52 &             4.1780 &    -0.573 &  724.235 & 102 &      & 0.06 4 & 0.04/0.22 3/5 & 0.04/0.24 4/5 & 0.10/0.25 3/4 &      &      & 0.14/0.28 4/4 & 0.06/0.14 3/5 \\
         5679.02 &             4.6520 &    -0.820 & 1106.291 & 156 &      &      & 0.04/0.12 5/5 & 0.03/0.30 4/5 & 0.08/0.26 5/5 &      &      & 0.05/0.07 5/5 & 0.04/0.11 4/4 \\
         5691.50 &             4.3010 &    -1.420 &  746.231 & 156 &      &      &      &      &      &      &      & 0.02/0.04 3/3 &      \\
         5696.09 &             4.5480 &    -1.720 &  965.279 & 102 &      &      &      &      & 0.01/0.28 3/4 &      &      & 0.03/0.07 3/3 &      \\
         5698.02 &             3.6400 &    -2.580 &  385.252 & 156 &      &      &      &      & 0.01/0.22 3/3 &      &      &      &      \\
         5701.54 &             2.5590 &    -2.160 &  361.237 & 102 &      & 0.06 4 & 0.05/0.28 4/5 &      & 0.10/0.28 3/3 &      &      & 0.09/0.11 4/4 & 0.06/0.16 4/5 \\
         5705.46 &             4.3010 &    -1.355 &  744.231 & 102 &      &      & 0.04/0.11 4/5 &      & 0.03/0.23 5/5 & 0.12/0.49 3/4 & 0.48 3 & 0.04/0.06 5/5 & 0.03/0.16 4/5 \\
         5731.76 &             4.2560 &    -1.200 &  727.232 & 156 &      &      & 0.02/0.15 5/5 & 0.02/0.15 4/5 & 0.09/0.38 4/5 &      & 0.58 3 & 0.03/0.06 5/5 & 0.03/0.07 4/5 \\
         5732.30 &             4.9910 &    -1.460 &  613.275 & 156 &      &      &      &      & 0.02/0.11 4/5 & 0.11/0.18 3/4 &      & 0.04/0.10 4/4 &      \\
         5741.85 &             4.2560 &    -1.672 &  725.232 & 102 &      &      & 0.03/0.17 5/6 &      & 0.05/0.25 6/6 & 0.07/0.23 3/4 & 0.62 3 & 0.03/0.06 6/6 & 0.05/0.17 4/5 \\
         5760.34 &             3.6420 &    -2.390 &  386.250 & 156 &      &      &      &      &      &      &      & 0.03/0.07 3/3 &      \\
         5775.08 &             4.2200 &    -1.297 &  720.231 & 102 &      &      &      &      & 0.07/0.34 5/6 & 0.04/0.39 3/5 &      &      & 0.05/0.07 5/5 \\
         5778.45 &             2.5880 &    -3.430 &  361.237 & 102 &      &      &      &      & 0.06/0.27 4/6 & 0.03/0.44 3/4 &      &      & 0.04/0.10 4/4 \\
         5784.66 &             3.3960 &    -2.532 &  796.244 & 102 &      &      &      &      & 0.05/0.42 3/3 &      &      &      &      \\
         5849.68 &             3.6940 &    -2.890 &  379.305 & 156 &      &      &      &      & 0.02/0.10 3/4 & 0.09/0.13 3/4 &      & 0.02/0.02 3/3 &      \\
         5853.15 &             1.4850 &    -5.180 &    0.000 & 156 &      &      &      &      & 0.02/0.16 3/4 &      &      & 0.03/0.05 3/3 &      \\
         5855.08 &             4.6080 &    -1.478 &  962.279 & 102 &      &      & 0.03/0.17 5/6 &      & 0.04/0.33 5/6 & 0.14/0.49 3/5 &      & 0.04/0.11 5/6 &      \\
         5858.78 &             4.2200 &    -2.160 &  786.278 & 156 &      &      &      &      & 0.01/0.09 3/4 & 0.02/0.18 3/3 &      & 0.02/0.08 3/3 &      \\
         5883.82 &             3.9600 &    -1.260 &  998.250 & 156 &      &      & 0.01/0.15 3/4 & 0.01/0.12 3/3 &      &      &      & 0.06/0.07 3/4 & 0.03/0.08 3/3 \\
         5902.47 &             4.5930 &    -1.710 &  227.252 & 156 &      &      &      &      & 0.06/0.19 3/4 &      &      & 0.11/0.26 3/3 &      \\
         5905.67 &             4.6520 &    -0.690 &  994.282 & 156 &      &      & 0.01/0.16 4/5 & 0.02/0.15 4/5 &      &      &      & 0.06/0.21 4/4 & 0.03/0.16 3/3 \\
         5927.79 &             4.6520 &    -0.990 &  984.281 & 156 &      &      & 0.02/0.09 3/4 &      & 0.05/0.16 3/4 & 0.14/0.46 3/4 & 0.39 3 & 0.04/0.05 3/4 & 0.01/0.08 3/3 \\
         5929.68 &             4.5480 &    -1.310 &  864.275 & 156 &      &      & 0.02/0.14 3/4 &      & 0.03/0.76 3/4 &      &      & 0.03/0.05 3/4 & 0.03/0.04 3/3 \\
         5930.18 &             4.6520 &    -0.230 &  983.281 & 187 &      & 0.07 3 & 0.04/0.19 3/5 & 0.01/0.13 3/5 & 0.06/0.17 3/5 &      &      & 0.14/0.19 3/4 & 0.02/0.05 4/4 \\
         5934.65 &             3.9280 &    -1.070 &  959.247 & 156 &      &      & 0.02/0.18 3/4 & 0.01/0.20 3/4 & 0.03/0.30 3/4 &      &      & 0.11/0.15 4/4 & 0.03/0.07 3/3 \\
         5956.69 &             0.8590 &    -4.553 &  227.252 & 102 &      & 0.04 4 &      &      &      &      &      & 0.03/0.10 4/5 & 0.03/0.16 4/4 \\
         6003.01 &             3.8810 &    -1.120 &  898.241 & 187 &      &      &      & 0.03/0.20 3/4 &      &      &      & 0.16/0.31 3/3 & 0.06/0.10 3/3 \\
         6012.21 &             2.2230 &    -4.038 &  309.270 & 102 &      &      &      &      &      &      &      & 0.03/0.05 3/4 & 0.04/0.12 3/4 \\
         6027.05 &             4.0760 &    -1.089 &  380.250 & 102 &      &      & 0.04/0.10 4/5 & 0.03/0.11 3/5 & 0.06/0.14 3/5 &      &      & 0.06/0.27 5/5 & 0.02/0.16 4/4 \\
         6065.48 &             2.6080 &    -1.470 &  354.234 & 102 & 0.03/0.05 4/4 & 0.05 4 &      & 0.04/0.29 3/5 &      &      &      &      & 0.09/0.34 3/4 \\
         6079.01 &             4.6520 &    -1.020 &  920.276 & 156 &      &      & 0.01/0.18 3/4 & 0.02/0.19 3/4 & 0.05/0.11 3/4 &      &      & 0.02/0.03 3/4 & 0.02/0.08 3/3 \\
         6093.64 &             4.6070 &    -1.400 &  866.274 & 156 &      &      & 0.01/0.06 3/4 &      & 0.02/0.07 3/4 & 0.05/0.20 3/3 &      & 0.03/0.04 3/4 &      \\
         6094.37 &             4.6520 &    -1.840 &  914.276 & 156 &      &      &      &      & 0.03/0.32 3/4 &      &      & 0.03/0.13 3/3 &      \\
         6096.66 &             3.9840 &    -1.830 &  963.250 & 156 &      &      & 0.03/0.23 4/5 & 0.02/0.08 4/4 & 0.04/0.18 4/5 &      &      & 0.04/0.14 5/5 & 0.03/0.08 3/4 \\
         6127.91 &             4.1430 &    -1.399 &    0.000 & 102 &      &      &      &      &      &      &      & 0.10/0.27 3/3 &      \\
         6136.99 &             2.1980 &    -2.941 &  280.265 & 102 &      & 0.02 3 & 0.05/0.37 3/4 &      &      &      &      & 0.05/0.12 3/4 & 0.02/0.07 3/3 \\
         6151.62 &             2.1760 &    -3.312 &  277.263 & 102 &      &      & 0.03/0.10 5/5 & 0.01/0.15 4/4 & 0.01/0.83 3/5 & 0.13/0.78 3/4 & 0.65 4 & 0.06/0.07 4/5 & 0.03/0.14 4/5 \\
         6165.36 &             4.1430 &    -1.473 &  380.250 & 102 &      &      & 0.02/0.18 4/5 & 0.01/0.07 3/5 & 0.05/0.64 4/5 & 0.15/0.32 3/4 &      & 0.04/0.06 5/5 & 0.04/0.16 4/5 \\
         6173.33 &             2.2230 &    -2.880 &  281.266 & 102 &      & 0.05 4 & 0.03/0.12 4/5 & 0.02/0.13 4/5 & 0.06/0.21 3/4 & 0.19/0.78 3/4 &      & 0.08/0.20 4/5 & 0.06/0.12 3/4 \\
         6187.99 &             3.9430 &    -1.620 &  903.244 & 156 &      &      & 0.03/0.22 5/6 & 0.02/0.20 5/6 & 0.05/0.65 5/6 &      &      & 0.04/0.21 6/6 & 0.03/0.08 4/5 \\
         6200.31 &             2.6080 &    -2.405 &  350.235 & 102 &      & 0.07 3 & 0.01/0.13 3/4 & 0.02/0.09 3/4 & 0.02/0.20 3/4 &      &      & 0.08/0.10 3/4 & 0.07/0.29 3/3 \\
         6219.28 &             2.1980 &    -2.434 &  278.264 & 102 &      & 0.02 4 & 0.02/0.40 3/5 & 0.04/0.26 3/5 & 0.02/0.28 3/3 &      & 0.10 3 & 0.09/0.17 3/4 & 0.05/0.07 4/4 \\
         6226.73 &             3.8830 &    -2.120 &  845.244 & 156 &      &      & 0.01/0.25 4/5 &      & 0.03/0.62 4/5 &      &      & 0.03/0.05 3/5 &      \\
         6240.65 &             2.2230 &    -3.203 &  301.272 & 102 &      &      &      &      &      &      &      & 0.09/0.38 3/3 &      \\
         6246.32 &             3.6020 &    -0.805 &  820.246 & 102 & 0.01/0.05 3/4 & 0.06 5 &      & 0.06/0.18 4/6 & 0.04/0.21 4/4 &      & 0.22 3 & 0.23/0.34 3/4 & 0.06/0.31 5/5 \\
         6252.56 &             2.4040 &    -1.727 &  326.245 & 102 & 0.02/0.05 5/5 & 0.04 5 & 0.07/0.26 4/6 & 0.07/0.14 3/6 & 0.09/0.26 3/4 &      & 0.37 3 & 0.17/0.19 3/4 & 0.11/0.35 3/5 \\
         6265.13 &             2.1760 &    -2.545 &  274.261 & 102 &      & 0.03 4 & 0.03/0.34 4/5 & 0.03/0.29 4/5 & 0.02/0.25 3/4 &      &      & 0.12/0.17 4/5 & 0.09/0.65 4/5 \\
         6270.22 &             2.8580 &    -2.536 &  350.249 & 102 &      &      & 0.03/0.29 4/5 & 0.03/0.17 3/5 & 0.06/0.76 4/5 &      &      & 0.04/0.05 4/5 & 0.02/0.27 4/5 \\
         6271.28 &             3.3320 &    -2.703 &  720.247 & 102 &      &      & 0.04/0.28 3/4 &      & 0.04/0.20 3/4 &      &      & 0.03/0.07 3/4 &      \\
         6297.79 &             2.2230 &    -2.702 &  278.264 & 102 &      & 0.04 3 &      & 0.04/0.15 3/4 &      &      &      & 0.15/0.18 4/4 & 0.08/0.19 3/4 \\
         6301.50 &             3.6540 &    -0.718 &    0.000 & 102 &      & 0.09 3 &      &      &      &      &      &      &      \\
         6315.81 &             4.0760 &    -1.610 &  410.250 & 156 &      &      & 0.02/0.35 3/4 & 0.04/0.21 3/3 & 0.05/0.18 3/4 &      &      & 0.05/0.10 4/4 & 0.02/0.53 3/4 \\
         6322.69 &             2.5880 &    -2.448 &  345.238 & 102 &      & 0.02 4 & 0.02/0.32 4/5 & 0.02/0.24 4/5 & 0.08/0.24 3/4 &      & 0.31 3 & 0.08/0.20 4/4 & 0.02/0.18 4/5 \\
         6335.33 &             2.1980 &    -2.177 &  275.261 & 102 & 0.03/0.09 3/3 & 0.05 4 & 0.03/0.14 3/5 & 0.05/0.15 3/5 & 0.08/0.28 3/3 & 0.23/0.27 3/3 &      & 0.18/0.22 3/3 & 0.12/0.28 4/4 \\
         6336.82 &             3.6860 &    -0.856 &  845.240 & 102 &      & 0.06 5 & 0.06/0.28 3/6 & 0.05/0.26 5/6 & 0.11/0.18 4/5 & 0.24/0.65 3/4 & 0.13 3 &      & 0.05/0.26 4/5 \\
         6393.60 &             2.4330 &    -1.504 &  326.246 & 102 & 0.04/0.07 5/5 & 0.06 5 & 0.05/0.23 3/5 & 0.07/0.45 4/5 & 0.12/0.24 3/4 &      & 0.51 3 &      & 0.14/0.18 3/4 \\
         6411.65 &             3.6540 &    -0.656 &  820.247 & 102 & 0.03/0.06 4/5 & 0.05 5 &      & 0.06/0.12 4/6 & 0.04/0.26 3/4 &      & 0.51 3 &      & 0.11/0.42 4/5 \\
         6430.85 &             2.1760 &    -1.976 &  271.257 & 102 & 0.02/0.05 5/5 & 0.05 5 & 0.03/0.45 3/6 & 0.05/0.13 3/6 & 0.10/0.25 3/4 &      & 0.51 3 & 0.18/0.29 3/4 &      \\
         6481.87 &             2.2790 &    -2.985 &  308.243 & 102 &      & 0.03 4 & 0.02/0.42 4/5 & 0.02/0.41 3/5 & 0.07/0.28 3/4 &      &      & 0.04/0.10 3/5 & 0.00/0.07 3/4 \\
         6494.98 &             2.4040 &    -1.256 &  321.247 & 102 & 0.03/0.08 3/3 & 0.04 3 &      &      &      &      &      &      &      \\
         6496.47 &             4.7950 &    -0.530 &  925.279 & 156 &      &      &      &      & 0.07/0.37 3/4 &      &      & 0.03/0.08 3/4 & 0.01/0.02 3/3 \\
         6498.94 &             0.9580 &    -4.688 &  226.253 & 102 &      & 0.06 4 &      &      &      &      &      & 0.11/0.11 4/5 & 0.04/0.23 3/4 \\
         6533.93 &             4.5580 &    -1.360 &  908.277 & 156 &      &      &      &      & 0.01/0.07 3/4 &      &      & 0.02/0.03 3/4 &      \\
         6574.23 &             0.9900 &    -5.013 &    0.000 & 102 &      &      &      &      &      &      &      & 0.08/0.13 3/4 & 0.04/0.16 3/4 \\
         6593.87 &             2.4330 &    -2.394 &  321.247 & 102 &      & 0.03 3 & 0.01/0.15 3/4 & 0.03/0.09 3/4 & 0.03/0.26 3/3 &      &      & 0.08/0.11 4/4 & 0.09/0.23 3/3 \\
         6597.56 &             4.7950 &    -0.970 &  893.276 & 156 &      &      & 0.01/0.18 3/4 & 0.02/0.14 3/4 &      &      &      & 0.02/0.19 3/4 & 0.04/0.05 3/3 \\
         6609.11 &             2.5590 &    -2.676 &  335.245 & 102 &      & 0.04 3 & 0.02/0.42 3/4 & 0.04/0.17 3/4 & 0.09/0.28 3/3 &      &      & 0.05/0.06 3/3 & 0.02/0.38 3/4 \\
         6627.54 &             4.5480 &    -1.580 &  754.209 & 156 &      &      &      &      &      &      &      & 0.02/0.13 3/3 &      \\
         6648.08 &             1.0110 &    -5.918 &  229.254 & 102 &      &      &      &      &      &      & 0.50 3 & 0.02/0.04 3/3 &      \\
         6699.14 &             4.5930 &    -2.101 &  297.273 & 102 &      &      &      &      & 0.05/0.14 4/5 &      &      & 0.04/0.07 3/4 &      \\
         6703.57 &             2.7580 &    -3.060 &  320.264 & 156 &      &      & 0.04/0.11 3/4 &      &      &      &      & 0.02/0.03 3/3 &      \\
         6713.74 &             4.7950 &    -1.500 &  857.272 & 156 &      &      &      &      & 0.01/0.44 3/4 &      &      & 0.04/0.05 3/3 &      \\
         6739.52 &             1.5570 &    -4.794 &  256.244 & 102 &      &      &      &      & 0.01/0.67 3/4 &      &      & 0.01/0.03 3/3 &      \\
         6750.15 &             2.4240 &    -2.604 &  335.241 & 102 &      & 0.03 3 & 0.03/0.14 3/4 &      &      &      & 1.09 3 & 0.07/0.07 3/4 & 0.03/0.05 3/3 \\
         6810.26 &             4.6070 &    -0.986 &  873.275 & 102 &      &      &      &      & 0.05/0.43 3/4 &      &      &      & 0.02/0.12 3/3 \\
\end{longtable}
\end{landscape}

}
\twocolumn
{
\small
\begin{landscape}
\centering
\begin{table*}
\caption{ List of ``golden'' Fe~II lines for various groups of stars (see text for definition of groups). \label{tab:golden_Fe2}} 
\begin{tabular}{rrrrrcccccc}
\hline\hline
$\lambda$ [\AA] & E [eV] & $\log gf$ &    Waals & Ref & MPD  (2) & MPG  (1) & FGD$^\dagger$ (15) & FGKG (7) & MG   (4) & KD$^\ddagger$ (1)\\
\hline
         4923.93 &             2.8910 &    -1.260 &  175.202 & 158 & 0.16/0.25 4/4 &      &      &      &      &      \\
         4993.36 &             2.8070 &    -3.684 &  172.220 & 166 &      & 0.04 3 & 0.00/0.21 3/4 & 0.13/0.64 3/4 &      & 0.29 4 \\
         5264.81 &             3.2300 &    -3.130 &  186.300 & 158 &      &      &      &      &      & 0.09 3 \\
         5325.55 &             3.2210 &    -3.160 &  179.252 & 158 &      &      &      &      &      & 0.08 3 \\
         5414.07 &             3.2210 &    -3.580 &  185.303 & 158 &      &      & 0.01/0.07 3/4 & 0.07/0.24 3/4 &      & 0.08 3 \\
         5425.26 &             3.1990 &    -3.220 &  178.255 & 158 &      & 0.14 3 & 0.01/0.14 4/5 & 0.09/0.35 4/5 & 0.17/0.50 3/5 & 0.11 5 \\
         5534.85 &             3.2450 &    -2.865 &  178.239 & 166 &      &      &      &      &      & 0.06 3 \\
         5991.38 &             3.1530 &    -3.647 &  172.221 & 166 &      &      & 0.01/0.12 3/4 &      &      & 0.04 3 \\
         6084.11 &             3.1990 &    -3.881 &  173.223 & 166 &      &      &      & 0.08/0.20 3/4 &      & 0.03 3 \\
         6247.56 &             3.8920 &    -2.435 &  186.272 & 166 &      &      &      &      &      & 0.04 3 \\
         6432.68 &             2.8910 &    -3.570 &  169.204 & 158 &      & 0.07 3 & 0.01/0.32 3/4 & 0.02/0.07 3/4 & 0.10/0.19 3/4 & 0.05 3 \\
         6456.38 &             3.9030 &    -2.185 &  185.276 & 166 &      & 0.14 4 & 0.03/0.19 4/5 & 0.04/0.38 4/5 & 0.14/0.24 4/4 & 0.03 5 \\
\hline\hline
\multicolumn{11}{l}{\parbox{1.00\textheight}{Column descriptions: $\lambda$: wavelength. $E_{\rm low}$: lower level energy. ``Waals'': parameters used to calculate line broadening due to collisions with neutral hydrogen; integer part: broadening cross-section at a velocity of 10$^4$~m s$^{-1}$ in atomic units, fractional part: velocity parameter \citep[see][]{BPM}; if zero, the Uns\"old approximation was used. ``Ref'': reference code for the $gf$-values (see below). The remaining columns are headed by a label for each group defined in the text, and the number of stars in parantheses. The columns give, for each group, the minimum and maximum standard deviations of the average line abundances, and the minimum and maximum number of abundances averaged for each line.}}\\
\multicolumn{11}{l}{\parbox{1.00\textheight}{Notes: $^\dagger$ 5414.07\AA\ was not used in \muCas, and 5991.38\AA\ not in HD~49933. $^\ddagger$ The column marks the lines used for \epsEri; of these, only 4993.36\AA\ and 6456.38\AA\ were used in 61~Cyg~A and Gmb~1830; only 4993.36\AA\ and 5425.26\AA\ in 61~Cyg~B.}}\\
\multicolumn{11}{l}{\parbox{1.00\textheight}{References: 158: \citet{2009AaA...497..611M}.  166: \citet{1998AaA...340..300R}. }}\\
\end{tabular}
\end{table*}
\end{landscape}

}

\end{document}